\documentclass{aastex}
\usepackage{spr-astr-addons}
\usepackage{url}

\usepackage{epsfig}
\usepackage{color}
\usepackage{subfigure}
\usepackage{multirow}
\usepackage{multicol}

\usepackage[breaklinks]{hyperref}
\usepackage{etoolbox}
\apptocmd{\thebibliography}{\raggedright}{}{}

\hypersetup{urlcolor=blue, colorlinks=true}
\usepackage{longtable}
\usepackage{amssymb}
\usepackage{lscape}
\usepackage{forloop}

\def\lsim{\mathrel{\rlap{\lower3.5pt\hbox{\hskip0.5pt$\sim$}}
    \raise0.5pt\hbox{$<$}}}
\def\gsim{~\rlap{$>$}{\lower 1.0ex\hbox{$\sim$}}}

\newcommand*\ud{\mathop{}\!\mathrm{d}}

\newcommand{\Swift}{\textit{Swift} }
\newcommand{\Luminosity}{erg s$^{-1}$ Hz$^{-1}$\,}
\newcommand{\Fluence}{erg cm$^{-2}$\,}
\newcommand{\Flux}{erg cm$^{-2}$ s$^{-1}$\,}

\newcommand{\emaila}{acruggeri@na.infn.it; capozziello@na.infn.it}

\begin{document}

\title{Possible Gamma-Ray Burst radio detections by the Square Kilometre Array. New perspectives}
\slugcomment{Possible GRB radio detections by the SKA. New perspectives}
\shorttitle{Possible GRB radio detections by the SKA. New perspectives}
\shortauthors{Ruggeri \& Capozziello}

\author{Alan Cosimo Ruggeri\altaffilmark{1,2,*}} 
\and
\author{Salvatore Capozziello\altaffilmark{1,2,3*}}

\altaffiltext{1}{Dipartimento di Fisica, Universit\`a di Napoli ``Federico II'', Compl. Univ. Monte S. Angelo, Via Cinthia 9, I - 80126, Napoli, Italy.}
\altaffiltext{2}{I.N.F.N., Sez. di Napoli, Compl. Univ. Monte S. Angelo, Edificio G, Via Cintia, 26 - 80126, Napoli, Italy.}
\altaffiltext{3}{Gran Sasso Science Institute (INFN), Viale F. Crispi, 7, I-67100, L'Aquila, Italy. }
\altaffiltext{*}{\emaila}


\begin{abstract}
The next generation interferometric radio telescope, the Square Kilometre  Array (SKA), which will be the most sensitive and largest radio telescope ever constructed, could greatly contribute to the detection, survey and characterization of Gamma Ray Bursts (GRBs). 
By the SKA, it will be possible to  perform the follow up of GRBs even for several months. This approach would be extremely useful to extend the Spectrum Energetic Distribution (SED) from the gamma  to the to radio band and would increase the number of radio detectable GRBs. 
In principle, the SKA could help to understand the physics of  GRBs by setting  constraints on  theoretical models. This goal could be achieved by taking into account multiple observations at different wavelengths in order to obtain a deeper insight of the sources. 
Here, we present an estimation of GRB radio detections, showing that the GRBs can  really be observed by the SKA. The approach  that we present consists in determining   blind detection rates derived by a very large sample consisting of merging several   GRB catalogues observed by current missions as \textit{Swift}, \textit{Fermi}, Agile and INTEGRAL and by previous missions as BeppoSAX, CGRO, GRANAT, HETE-2, Ulysses and Wind. The final catalogue counts 7516 distinct sources. We compute the fraction of GRBs that could be observed by the SKA at high and low frequencies, above its observable sky. Considering the planned SKA sensitivity and through an extrapolation based on previous works and observations, we  deduce the minimum fluence in the range  15\,-\,150 keV.  This is the energy interval where a GRB should emit to be detectable  in the radio band by the SKA. Results seem consistent with observational capabilities.

\end{abstract}



\section{Introduction} \label{Introduction}
Due to their high fluence,   between $10^{-7}$ and $10^{-5}$ erg/cm$^2$, and to their huge isotropic energies, between  $\sim 10^{48}$ - $10^{54}$ erg emitted in a very short time, Gamma Ray Bursts (GRBs) are the most violent and energetic astrophysical phenomena currently known  in the Universe. \\
Discovered in 1963 but announced only in the 1974 \citep{annuncioGRB}, the study of GRB  is  continuing  and improving. Although GRBs are the brightest sources in the Universe  and studied  for long time, there are several  unclear aspects  that has to be understood.
For example,  several  models  try to explain the physics of these phenomena, but none can be considered   final and  self-consistent. One of the principal issues is that GRB spectra are very different from each other and this fact makes them very difficult to study and classify: in general,  different spectra correspond to various combinations of parameters.  So far, the central engine of these sources is still debated (see, e.g., \cite{FWK00,Da03,Mes06}), because there is no single theoretical model capable of explaining in a comprehensive way all the observations. \\
In general, they are cosmological objects and may occur at any point of the Universe, namely at different redshift, and, for this reason,   strong selection effects  plays an important rule in the discussions about  the physics of GRBs. Currently, the farthest GRBs have been detected  at $z = 8.1$ \citep{Salvaterra2009} (spectroscopic redshift) and $z = 9.4$ \citep{GRBrecord} (photometric redshift). Using  them as distance indicators  is crucial  issue in order to probe the Hubble flow  up to  early epochs. \par 
In general, GRBs are peculiar sources and because of the variety of light curves and  spectra, many efforts are concentrating on a possible standardization. These attempts are principally focused on high energies (i.e., $\gamma$\,-\,rays, X\,-\,rays), even though some recent studies are  extending the frequency range towards the low energies. The general aim is to discover a standard behavior for a specific class  of GRBs  presenting  defined  characteristics and features. \\ 
 In other words, it is necessary to observe GRBs in several spectral bands and  consider their complete envelopes in a large energy range. Studying in detail a large  number of Spectral Energy Distributions (SEDs) could allow  to highlight the emission process in a   large spectral band. In this sense, a crucial breakthrough  has been   achieved by the launch of the {\it Swift} satellite in 2004. The \textit{Burst Alert Telescope} (BAT) in 15\,-\,150 keV energy range, the \textit{X\,-\,Ray Telescope} (XRT) in  0.3\,-\,10 keV and the \textit{Ultra-Violet/Optical Telescope} (UVOT) in 170\,-\,650 nm make up the payload of \textit{Swift} and allow a rapid follow-up of the afterglows in different wavelengths. These facilities give  a better coverage of the GRB light curve than the previous satellite missions. By the instrumentation for the X and UV counterparts, \textit{Swift} allows a rapid localization of GRBs and several efforts have been dedicated to trace afterglow light curves at different wavelengths \citep{N06,K11,O09}.\\
By using large catalogues, comparative studies among optical and X\,-\,ray light curves allow to fix constraints on GRB theoretical models (see, e.g., \cite{O09}, \cite{S11} for the standard fireball model) and to carry out correlations among optical and X\,-\,ray properties, e.g. between fluence and brightness \citep{G08, N09, K11}. \par
Regarding to the radio band, the GRB light curve could be tracked for hundreds of days after the $\gamma$ \,-\,ray onset, but fluxes are very faint at these  frequencies and only a few of current instruments can detect them.
This paper, without claiming for completeness, is a discussion on interferometric radio observations since  these could improve the detection, the surveys and the characterization of GRBs at frequencies which are not usually investigated. Our aim  is to show that these observations are a realistic option.
 In general, the analysis of the GRB afterglow light curves, at different wavelengths, allows to investigate the physics of these fascinating objects probing how the blast wave generated by the burst propagates in the circumburst medium. At the present state of the art, it is crucial to boost the multi-wavelength study of the afterglow by radio observations and, in particular, by the Square Kilometre Array (\href{http://www.skatelescope.org/}{SKA}\footnote{http://www.skatelescope.org/}). \\
Although satellite observations are the primary step for the GRB detection, they are not and cannot be the only observational channel. Satellite missions are very expensive, have limited lifetimes and something could go amiss (e.g., malfunctioning, mission delays, wrong orbiting, etc). This means  that  satellite GRB investigations should be supported  with ground-based  telescopes, promoting,  finally,  research towards radio frequencies. It is worth noticing  that radio band is not affected by radiation extinction, contrary to higher frequencies. An accurate calorimetry in the  radio band  for  well-detectable GRBs would be possible in principle. In addition, radio observations can give  substantial indications   of the inverse Compton effect, since only radio frequencies can probe the density of the interstellar medium. In other words,  redshift measurements of the host galaxies can be obtained observing the hydrogen spin-flip at 1.4 GHz.  These observations  give  fundamental information for  GRBs guested in galaxies. \par
The layout paper is  as follows: in Sec. \ref{importance of radio} we discuss GRBs in radio band; in Sec. \ref{GRBs above SKA} the SKA telescope is shortly introduced, considering, in particular,  its instrumental sensitivities. Specifically, in  subsections (Sec. \ref{Collection of GRB catalogues}, \ref{sec: The sky above the SKA and its shadow cone}, \ref{Sec: A suggestion for a Spectral Energy Distribution from radio to gamma band}, \ref{Sec: The SKA sensitivity and the radio GRB-detection probability}, \ref{sec: Serendipitous detection rate probability for the SKA}), we calculate the minimum fluence between 15\,-\,150 keV to detect a GRB by the SKA introducing a first simple SED from gamma range to radio band,  give some estimations of the GRB detection rates for the SKA at high and low frequencies. We  use a catalogue assembled by  collecting  several previous catalogues and obtain a large list with 7516 GRBs. Discussions and conclusions are given  in Sec. \ref{sec: Discussions and conclusions}.
\section{GRBs and their radio emissions}\label{importance of radio}
Notwithstanding the peculiarities of single GRBs, both optical and X\,-\,ray afterglows share the feature of an exponential decline with  time $t$ since the burst. Their light curves cannot be followed for more than few days, with exception of some particular long GRBs such as GRB\,980425 or GRB\,060218 \citep{Dainotti2007, Bernardini2008}. On the other hand, radio afterglows have a shallower fall off, extending the light-curve decay up to hundredths of days since the burst. \par
Because of their faint radio emission, these observations are quite difficult and the number of GRBs with radio detections is small if compared to those observed in both optical and X\,-\,ray wavelengths. In fact, since January 1997 to April 2011, the radio list counts 304 GRBs observed and collected by \cite{Chandra2012}. The current radio detection rate is about $30\%$ \citep{Chandra2012}, but it might be limited by the sensitivity of the current radio telescopes and interferometers. The GRB radio emissions have been observed between $0.1$ and $300$ days, within a frequency range between 600 MHz and 660 GHz. Radio detections have typical flux densities of $150-200$ $\mu$Jy, while the $3 \sigma$ upper limits are at the level of $100-150$ $\mu$Jy.  \par
One year after the work by  \cite{Chandra2012}, the radio \textit{non-detections} and \textit{detections} identified in their sample were thought as belonging to two distinct populations by \cite{Hancock2013}. They named the radio non-detections as \textit{radio-faint} (RF) and the radio detections as \textit{radio-bright} (RB), thus dividing the GRBs into  two radio groups (or populations). The reason of this classification is that the detection rates at radio wavelengths are significantly lower than those at X-rays and optical frequency, and that this difference could not depend on the  instrumental sensitivity, because the means between RB and RF GRBs differ by up to three orders of magnitude, perhaps due to intrinsic differences. Nevertheless, the limits between the  two populations are not sharp, in particular  during  the first two weeks after  the burst, while in later times, fainter radio afterglows  tend to dim. The correct fraction between RF and RB is $\sim 30$ - $40\%$ and $\sim 60$ - $70\%$, respectively. With an additional investigation, they show that the two populations follow particular distributions at different wavelengths. Even though the $20$ - $40\%$ of RF GRBs could be bright, the difference between the group seems to be effective  and confirmed at other frequencies. \\
Where Chandra \& Frail do not give  final response about radio non-detections, apart from an instrumental sensitive limit, Hancock et al. give a theoretical interpretation, providing a good estimate about the flux density of radio emissions of RF GRBs, in a good agreement with the previous radio observational data. Their estimation is used in Sec. \ref{Sec: The SKA sensitivity and the radio GRB-detection probability}.
\section{GRB Statistics within the SKA field of view} \label{GRBs above SKA} 
%
%
The next radio telescope able to observe GRB radio afterglows will be the SKA. It will be the largest and most sensitive radio telescope in the world, composed by  three antenna designs which will be able to observe in a large frequency range, from 50 MHz to $\sim 20$ GHz. The ranges of these three kinds of antennas will be for low, middle and high frequency. Respectively, the Low-Frequency Aperture Array (LFAA, or SKA-Low) will cover the lowest frequency band, from 50 MHz up to 350 MHz; the SKA Mid Frequency Aperture Array (MFAA) is scheduled in a later phase of the construction and will cover from 400 MHz upwards; finally, 3000 single dish antennas are planned to observe from around 350 MHz to 20 GHz, (for details see \cite{Baseline.Design}). \\
By using interferometry, the SKA will have a combined collecting area of approximately one square kilometre, with its $15$ m diameter dishes, LFAA and MFAA antennas. Regarding the high frequency, receivers of the dish are planned to be of two types, i.e., Single Pixel Feeds (SPFs) in South Africa and Phased Array Feeds (PAFs) in Australia, with different reception characteristics. For this reason, two different names have been chosen to distinguish these dishes, i.e., SKA-Mid with the SPFs and SKA-Survey with the PAFs. 
Its field of view (FoV) will be equal to 200 square degrees between 70 and 300 MHz, from 1 to 200 square degrees between 300 MHz and 1 GHz and 1 square degree maximum between 1 and 10 GHz.  The SKA telescope will have sensitivities of 3.72, 2.06 and 0.72 $\mu$Jy-hr$^{-1/2}$, respectively with SKA-Survey, SKA-Low and SKA-Mid \citep{Baseline.Design}. To give a more precise idea about SPF bands, they will be five between 0.35\,-\,1.05 GHz, 0.95\,-\,1.76 GHz, 1.65\,-\,3.05 GHz, 2.80\,-\,5.18 GHz and 4.60\,-\,13.8 GHz, but only the bands 1 and 2 will be populed during the first phase of construction. Currently, the sensitivities $[\begin{large}\eta                                                                                                                                                                                                                                                                                                                                                                                                                                                                                                                                                                                                                                                                          \end{large}_{\rm{feed}} \,  A_{\rm{phy}}/T_{\rm{sys}}]$ at the different frequencies will respectively be 4.2, 7.0, 7.0, 6.4, 6.2 m$^2$/K. In general, the minimum detectable flux density is 
\begin{equation}
 \Delta S \approx 2 k_B \, \frac{T_{\rm{sys}}}{\begin{large}\eta \end{large}_{\rm{feed}} \, A_{\rm{phy}}} \, \frac{1}{\sqrt{2 \, \Delta \nu \, \Delta t}} \, \frac{1}{\sqrt{N (N -1)}} \,,
\end{equation}
where $T_{\rm{sys}}$ is the system temperature, $A_{\rm{phy}}$ is the physical area of the dish, $\begin{large}\eta                                                                                                      \end{large}_{\rm{feed}}$ is the antenna efficiency \citep{Wrobel.Walker}. These factors depend on the dish structure. Here $k_B$ is the Boltzmann constant,  $\Delta t$, $\Delta \nu$ and $N$ are respectively the integral observational time, the bandwidth and the number of antennas used for the observation. To give an idea of the minimum detectable flux densities at those five frequencies, we can take into account 50 antennas, 30 minutes in observations in continuum band and bandwidths of 700, 808, 1403, 2380 and 9200 MHz. The results are respectively: $8.36 \cdot 10^{-1}$, $4.67 \cdot 10^{-1}$, $3.54 \cdot 10^{-1}$, $2.98 \cdot 10^{-1}$ and $1.56 \cdot 10^{-1}$ $\mu$Jy. \\ 
Thanks to the interferometric technique with a baseline of about 3000 km and a planned angular resolution of 0.1 arcsec at 1 GHz, the SKA will be able to operate surveys of the sky at a rate faster than any survey telescope that has ever existed. It will perform continuous surveys looking at vast swathes of the radio sky from the southern hemisphere and part of the northern one, making a detailed map as the thousands of telescopes work in unison. The continuity of the surveys will give the possibility to observe transients, such as GRBs. It will allow to obtain their light curves in radio frequencies, currently detectable with difficulty, and only if very bright in this band. Instrumental synergies of gamma satellite missions and radio telescopes, e.g., \textit{Swift} and AMI (Arcminute Microkelvin Imager) \citep{Staley2013, Anderson2014}, could give an excellent chance to the SKA monitor program. In fact, high frequency dishes could be on-target already a few minutes after the  explosion, thus obtaining early-time light curves of GRBs. Furthermore, thanks to the FoV and range of the LFAA, one will have the opportunity to observe transients whose GRBs either elude gamma-ray satellites, or are impossible to detect at high energies such as orphan afterglows \citep{Ghirlanda2014rivista}. In these cases no gamma-alert is needed. \\
For more information, we suggest to see the SKA science books too (e.g. \cite{ItalianWB} and the international white book\footnote{\href{http://arxiv.org/find/all/1/all:+EXACT+Science_with_the_Square_Kilometer_Array/0/1/0/all/0/1}{http://arxiv.org/find/all/1/all:+EXACT+\\Science\_with\_the\_Square\_Kilometer\_Array/0/1/0/all/0/1}}). \par
In the next subsections, we discuss  radio observations and statistics following two cases. The first case concerns the usual method for radio GRB detections, i.e. satellite detections and then radio observations. The second case concerns the  serendipitous radio detection. Regarding the first case, it is important to highlight that we use  satellite observations for our statistics. Since we have to consider only obtained gamma detections, an ``instrumental probability'' will be more plausible instead of a (theoretical) enlarged detection related to all the possible detections. In other words, we firstly consider only the probability for an ensemble of already \textit{detected} GRBs and, secondly, the likelihood concerning  all the  \textit{detectable} GRBs. 
\subsection{The sample selection} \label{Collection of GRB catalogues}
As mentioned before, to investigate the potentialities of the SKA related to the GRB detections in radio band, we need a good instrumental probability, so that we present here a useful repository of all GRBs taken by several catalogues. The general approach  pursued to obtain this large catalogue is explained in this subsection. Other details are given in \hyperlink{sec: Appendix A}{Appendix A}. This repository, associated to radio considerations, has allowed us to find a final number of sources that could  possibly be observed by the SKA after satellite detections.  In addition, the complete table reported on line will make faster the   searches for given GRBs.  \\
The following satellite missions and the associated catalogues have been taken into account: Agile by \cite{Galli2013, Pal'shin2013, Longo2012, Hurley2013}, BeppoSAX by \cite{BeppoSAX}, CGRO by \cite{BATSE.Stern, BATSE.Kommers, CGRO.COMPTEL} and the \href{http://heasarc.gsfc.nasa.gov/W3Browse/cgro/batsegrb.html}{\textit{batsegrb} catalogue} \citep{BATSE4B.Meegan, BATSE3B.Meegan} in HEASARC web site\footnote{http://heasarc.gsfc.nasa.gov/W3Browse/cgro/batsegrb.html. \\
However, there is another useful web site at \\ \href{http://gammaray.msfc.nasa.gov/batse/grb/catalog/current/}{http://gammaray.msfc.nasa.gov/batse/grb/catalog/current/}.}, \textit{Fermi} with \cite{Fermi, Fermi2.vonKienlin} and the \href{http://heasarc.gsfc.nasa.gov/W3Browse/fermi/fermigbrst.html}{\textit{fermigbrst} catalogue} in HEASARC archive\footnote{http://heasarc.gsfc.nasa.gov/W3Browse/fermi/fermigbrst.html} \citep{Fermi.HEASARC.Gruber, Fermi2.vonKienlin, Fermi.HEASARC.Goldstein, Fermi}, GRANAT, HETE-2, INTEGRAL by \cite{Mineav2014, Bosnjak2014, Mereghetti2013}, Konus/Wind by  \cite{Pal'shin2013}, \textit{Swift} by \cite{Swift} and the \href{http://swift.gsfc.nasa.gov/archive/grb_table}{``\textit{Swift} GRB Table and Lookup''}\footnote{http://swift.gsfc.nasa.gov/archive/grb\_table}; ULYSSES by  \cite{ULYSSES}. By cross-correlating of different catalogues, we have created a list composed by 7516 distinct GRBs shown in Table \ref{Table: all GRBs} (here only a sample, the complete table is presented as on line material). Considering only GRBs that have coordinates in various catalogues, coincidences have been found in order to identify equal sources and discard duplicates. \par
Table 2 in \cite{ULYSSES} has 218 GRBs, but, as mentioned in the associated paper, they are all coincident with 218 GRBs triggered by CGRO/BATSE. \par
The catalogues in \cite{BATSE.Stern} and in \cite{BATSE.Kommers} (both focused on BATSE receiver on-board CGRO) have been considered. The former counts 3906 GRBs (2068 triggered and 1838 non-triggered), whereas the second one contains 873 (only non-triggered). Precisely, 725 non-triggered plus 15 triggered elements out of (Kommers') 873 objects were considered by Stern like coincident GRBs in his catalogue. 133 GRBs by Kommers plus 1838 by Stern have been taken to make up a ``BATSE non-triggered catalogue'' (BNT) with 1971 GRBs in total. Subsequently, using the \textit{batsegrb} catalogue, and merging this with the previous 2068 GRBs in the Stern catalogue, the ``BATSE triggered catalogue'' (BT) with 2703 elements has been obtained. Finally, 4674 GRBs detected by BATSE/CGRO instrument have been selected. Apart from BATSE catalogues, also table 8 in \cite{CGRO.COMPTEL} has been consulted, for the COMPTEL instrument. In that table, the precise detection time lacks, but thanks to the reported days and coordinates,  it has been possible to establish that 29 out of 31 sources were present in BT catalogue, 1 out of 31 source was in BNT catalogue, and the last one had a coincidence with BT and GRANAT detections. \par
By considering Table 2 in \cite{BeppoSAX}, the catalogue counts 873 GRBs with their own coordinates. For not  considering  the same source between the BeppoSAX and CGRO catalogues, we assume two GRBs equal if their detection intervals are $\leq 0.005$ days (that is $\sim 7$ minutes) and if they are within an angle of $146^\circ$ (both RA and DEC)\footnote{The estimate for maximum location error is $63.7^\circ$ in \cite{BATSE.Stern} for BATSE; the maximum location error is $83^\circ$ in \cite{BeppoSAX} for BeppoSAX. However, the most discriminating factor is generally the time, indeed changing angular range modifies their overlapping of just a few elements.}. This solution is ad hoc for our tasks, but a similar method was used by Stern matching between non-triggered sources in his table and sources in Kommers' table. Finally, 421 (340 of which are in BT, while the remaining 81 are in BNT) are common sources between BeppoSAX and CGRO. A similar comparison has been done with other missions working in the same time-frame of BeppoSAX. \par
For HETE-2 mission, the following catalogues have been consulted: \href{http://heasarc.gsfc.nasa.gov/W3Browse/hete-2/hete2grb.html}{``\textit{hete2grb}: HETE-2 Gamma-Ray Bursts''}\footnote{http://heasarc.gsfc.nasa.gov/W3Browse/hete-2/hete2grb.html} by \href{http://space.mit.edu/HETE/Bursts/}{MIT}\footnote{http://space.mit.edu/HETE/Bursts/} (Massachusetts Institute of Technology) and \href{http://heasarc.gsfc.nasa.gov/W3Browse/hete-2/hete2gcn.html}{``\textit{hete2gcn}: HETE-2 GCN Triggers Catalog''}\footnote{http://heasarc.gsfc.nasa.gov/W3Browse/hete-2/hete2gcn.html}. 
In these catalogues, because not every GRB has coordinates, we have cross-correlated the 84 GRBs of the first catalogue with the 1235 rows of the second one, thus obtaining 71 GRBs in common, associating them to their respective coordinates. No GRB is in common between HETE-2 and BeppoSAX, Konus/Wind, \textit{Swift}, only 1 with INTEGRAL is in common.\par
Then, we have considered the current missions, \textit{Swift} and \textit{Fermi}, using tables in \cite{Swift, Fermi, Fermi2.vonKienlin} and the HEASARC archive. Until 12th of May 2014, \textit{Fermi} satellite has reported 1359 GRBs and \textit{Swift} has reported 869\footnote{I report that in \cite{Swift} there are 3 elements more with respect to the web table in ``\textit{Swift} GRB Table and Lookup'', but they have not been counted.}. As previously, two detections are considered equal if they lie in a time range of 0.005 days and an angular range\footnote{The most position error (90\% error radius) in \cite{Swift} is of $6.4'$; the maximum positional uncertainty in \textit{Fermi} catalogue is $45.7^\circ$.} of $45^\circ$. We find 193 common sources between \textit{Fermi} and \textit{Swift}, but all coincidences with other missions are reported in Table \ref{Tab: Missions}. \par
Unfortunately, contrary to \textit{Fermi}, \textit{Swift} or CGRO, a large unique catalogue dedicated to gamma-ray bursts does not exist for the Agile mission, thus we have consulted the above-mentioned catalogues, and the \href{http://swift.gsfc.nasa.gov/archive/grb_table/}{\textit{Swift} web site}. We have always taken into account GRBs with their respective coordinates, thus obtaining the ``Agile list'' which collects 89 sources. Sources collected in literature are the results of triangulations with various instruments (e.g., Konus/Wind, INTEGRAL, HETE-2, Suzaku, etc) where the Agile contribution was used. The GRBs of Table 2 presented in \cite{Hurley2013} are InterPlanetary Network (IPN) identifications and the \textit{Fermi}/GMB was used for some of these triangulations. In those cases, they have already been included in \textit{Fermi} catalogues, cited here, and hence excluded from this final Agile list. \par
Konus/Wind, GRANAT and INTEGRAL  have the same issue as Agile. In fact, we have matched various catalogues (i.e., \cite{Mineav2014}, \cite{Bosnjak2014}, \cite{Mereghetti2013}, \cite{Pal'shin2013}), in order to identify single different objects with respective coordinates, and to have a reasonable list for each of these missions. For this purpose, the \href{http://simbad.u-strasbg.fr/simbad/}{SIMBAD Astronomical Database} and HEASARC database have also been used. From the latter, the \href{http://heasarc.gsfc.nasa.gov/W3Browse/gamma-ray-bursts/grbcat.html}{\textit{grbcat}}\footnote{http://heasarc.gsfc.nasa.gov/W3Browse/gamma-ray-bursts/grbcat.html} catalogue and the \href{http://heasarc.gsfc.nasa.gov/W3Browse/gamma-ray-bursts/phebus.html}{\textit{phebus}}\footnote{http://heasarc.gsfc.nasa.gov/W3Browse/gamma-ray-bursts/phebus.html} catalogue \citep{GRANAT.HEASARC1994, GRANAT.HEASARC1995, GRANAT.HEASARC1998, GRANAT.HEASARC2002} (this last in particular way for GRANAT) have been consulted. In this case, if the sources were without coordinates, we searched in SIMBAD or in \textit{grbcat} catalog. Obviously, great cure was taken for not reporting duplicates among catalogues of the same satellite. We have matched these last catalogues with previous ones by GRB names and detection time. \par
\begin{figure*}
\includegraphics[width=17 cm]{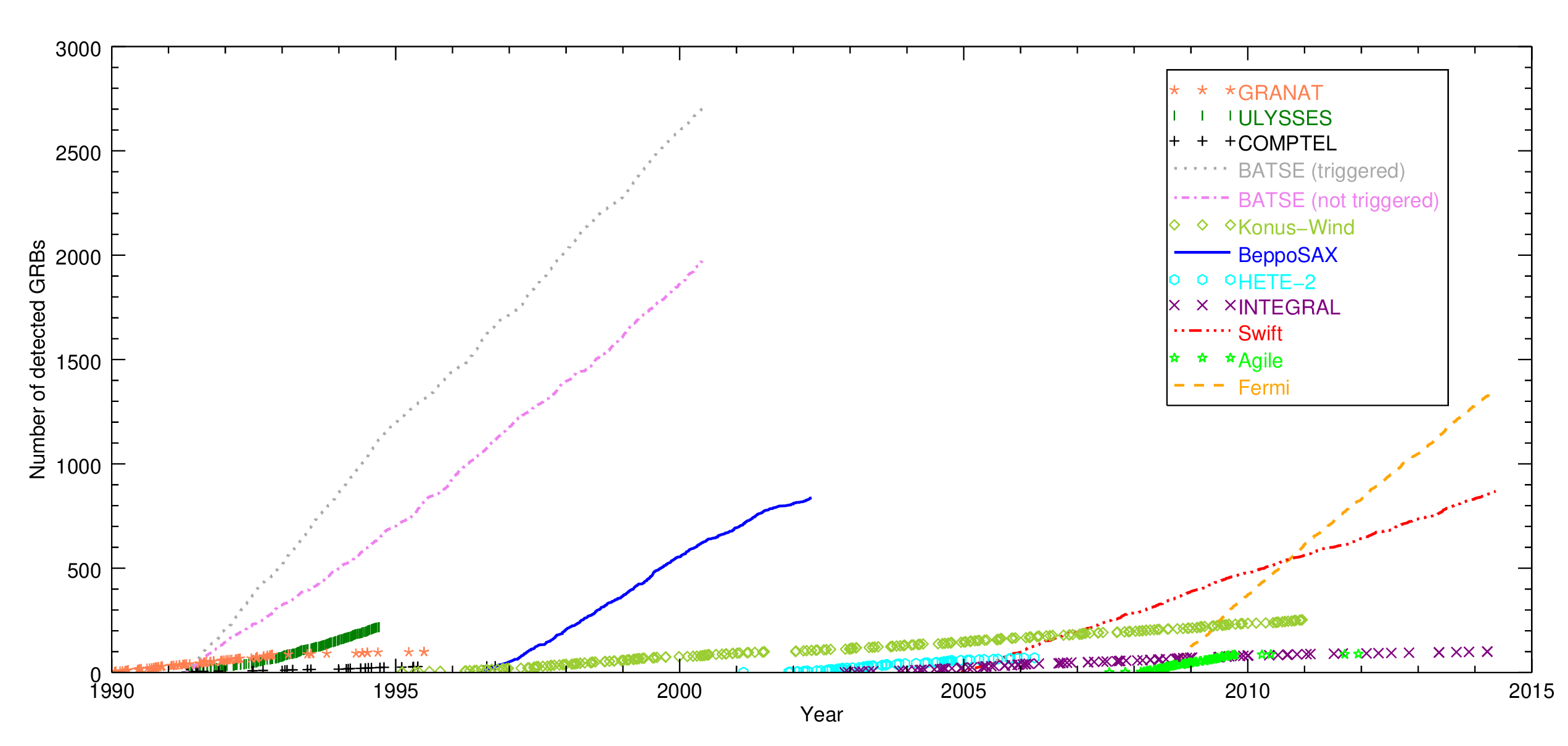}
\caption{In this figure, different missions, instruments and catalogues are considered. We report GRBs observed by GRANAT, Ulysses, CGRO/BATSE, CGRO/COMPTON, Konus/Wind, BeppoSAX, HETE-2, INTEGRAL, \textit{Swift}, Agile and \textit{Fermi}. We also plot the non-triggered GRBs by CGRO/BATSE. In the Y-axis the (progressive) number of GRBs detected by a mission/instrument is reported; in the X-axis the time of the detection expressed in years. Detections since 18th January 1990 done by GRANAT until 12th May 2014 done by \textit{Fermi} and \textit{Swift} are here plotted.}
 \label{fig: TimePlot}
\end{figure*}
\begin{figure*}
\includegraphics[width=17 cm]{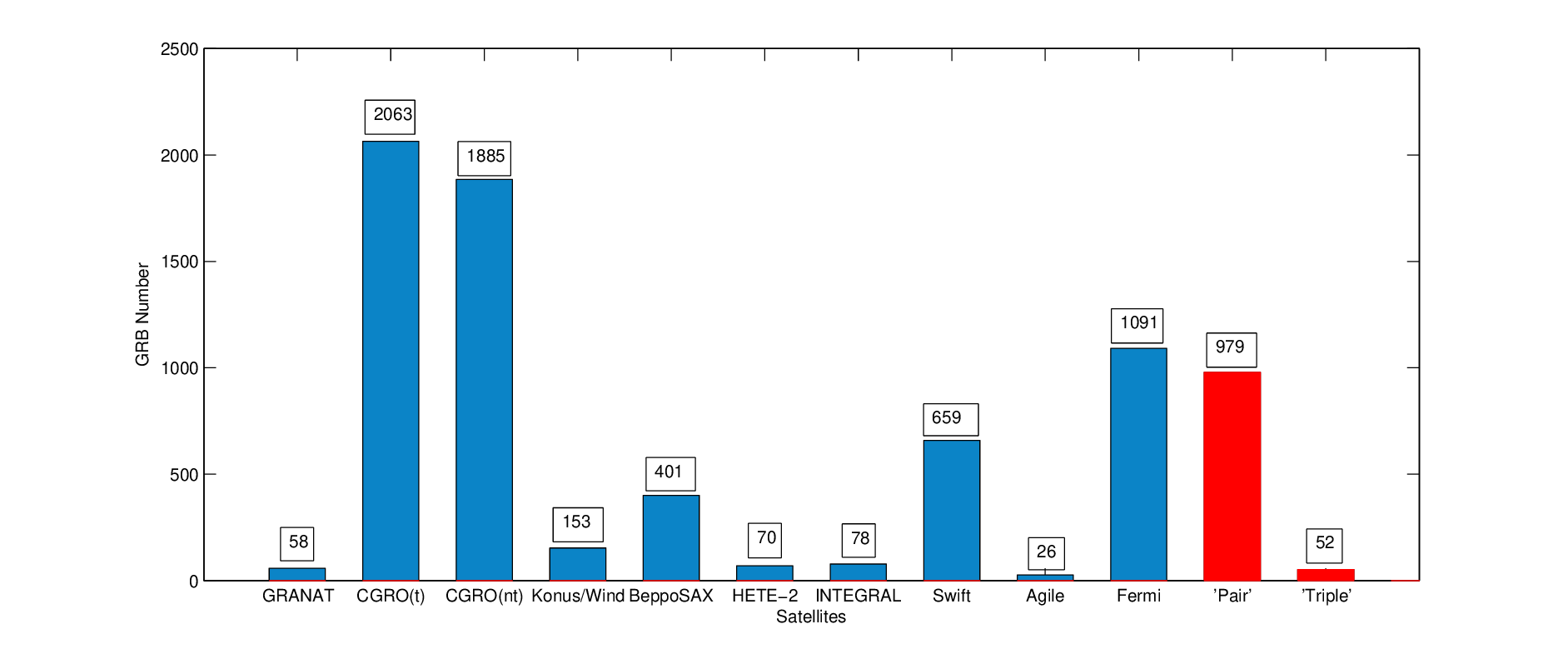}
\caption{In this histogram, 7516 distinct GRBs are allocated in different columns. The reported satellite detections occurred since 18$^{\rm{th}}$ January 1990 to 12$^{\rm{th}}$ May 2014, by GRANAT, Ulysses, CGRO, Konus/Wind, BeppoSAX, HETE-2, INTEGRAL, \textit{Switf}, Agile and \textit{Fermi} missions. As it is possible to see, we have considered single events for different missions/instruments, as well as same GRBs observed simultaneously by two or three satellite missions (or instruments) in the last two columns. See also Table \ref{Tab: Missions}. }
 \label{fig: istogramma}
\end{figure*}

Distributions of GRBs are presented in figure \ref{fig: istogramma}, where the histogram shows 12 different columns divided into three groups. The first ten columns report objects which have been detected by only one mission or one instrument, the eleventh column where same objects have been detected by two instruments (the ``couples'' column), finally, the last column where GRBs have been detected by three instruments simultaneously (label as ``triples''). %
Precisely, Table \ref{Tab: Missions} shows all the combinations, also distinguishing the triggered and non-triggered GRBs detected by BATSE. \\
The overlapping missions can be  easily  seen in Figure \ref{fig: TimePlot}, where detections are plotted in function of the time. The trigger time has been used to do this plot, which is conventionally written as ``year, month, day, universal time'' (YYYY MM DD hh:mm:ss). We wrote a routine that works as a counter, translating dates and hours in progressive numbers starting at the first gamma detection (18th January 1990) until the last considered for our estimations (12th May 2014). 
Table \ref{Tab: angular coefficients} lists an approximate detection-rate for each fitted line. For this last point, it is worth stressing that several  missions were not exclusively dedicated for GRB exploration, so those rates are only indicative; in fact some tables found in literature are not complete for each mission and they report only partial GRB detections. However, the most plausible detection-rates are CGRO, BeppoSAX, \textit{Swift} and \textit{Fermi}. 
\begin{table}
 \caption{Number of GRBs detected by the different  used catalogues and their combinations.\newline 
 {\small GRA = GRANAT; BT = BATSE triggered; BNT = BATSE non-triggered; COMP = COMPTEL; Uly = Ulysses; KW = Knous/Wind; BeS = BeppoSAX; Agi = Agile; Fer = \textit{Fermi}; INT = INTEGRAL; HET = HETE-2; Swi = \textit{Swift}.}}
 \begin{tabular}{l c}
   \hline
   \hline
      Mission & GRB Number \\
   \hline
      GRA          &   58  \\
      GRA-BT       &   26  \\
      GRA-BT-COMP  &   1   \\
      GRA-BT-Uly   &   14  \\
      GRA-BNT      &   1   \\
      \hline
      KW           &  153  \\
      KW-BT        &  26   \\
      KW-BT-BeS    &  18   \\
      KW-Agi       &  4    \\
      KW-Agi-Fer   &  7    \\
      KW-BNT       &  3    \\
      KW-BeS       &  15   \\
      KW-BNT-BeS   &  2    \\
      KW-INT       &  1    \\
      KW-Fer       &  22   \\
      \hline
      HET          &  70   \\
      \hline
      INT          &  78   \\
      INT-HET      &  1    \\
      INT-Swi      &  6    \\
      INT-Fer      &  14   \\
      INT-Fer-Agi  &  1    \\
      \hline
      Agi          &  26   \\
      Agi-Fer      &  31   \\
      Agi-Swi      &  11   \\
      Agi-Fer-Swi  &  9    \\
      \hline
      BT           &  2063 \\
      BT-Uly       &  204  \\
      BT-COMP      &  29   \\
      BT-BeS       &  322  \\
      \hline
      BNT          &  1885 \\
      BNT-COMP     &   1   \\
      BNT-BeS      &   79  \\
      \hline
      BeS          &   401 \\
      \hline
      Swi          &   659 \\
      Swi-Fer      &  184  \\
      \hline
      Fer          &  1091 \\
      
      \hline \hline
      Total        & 7516  \\
      \hline \hline
      
 \end{tabular}

 \label{Tab: Missions}

\end{table}
\subsection{The sky above the SKA and its shadow cone} \label{sec: The sky above the SKA and its shadow cone}
\begin{figure*}
  \centering
\fbox{\includegraphics[width=\columnwidth]{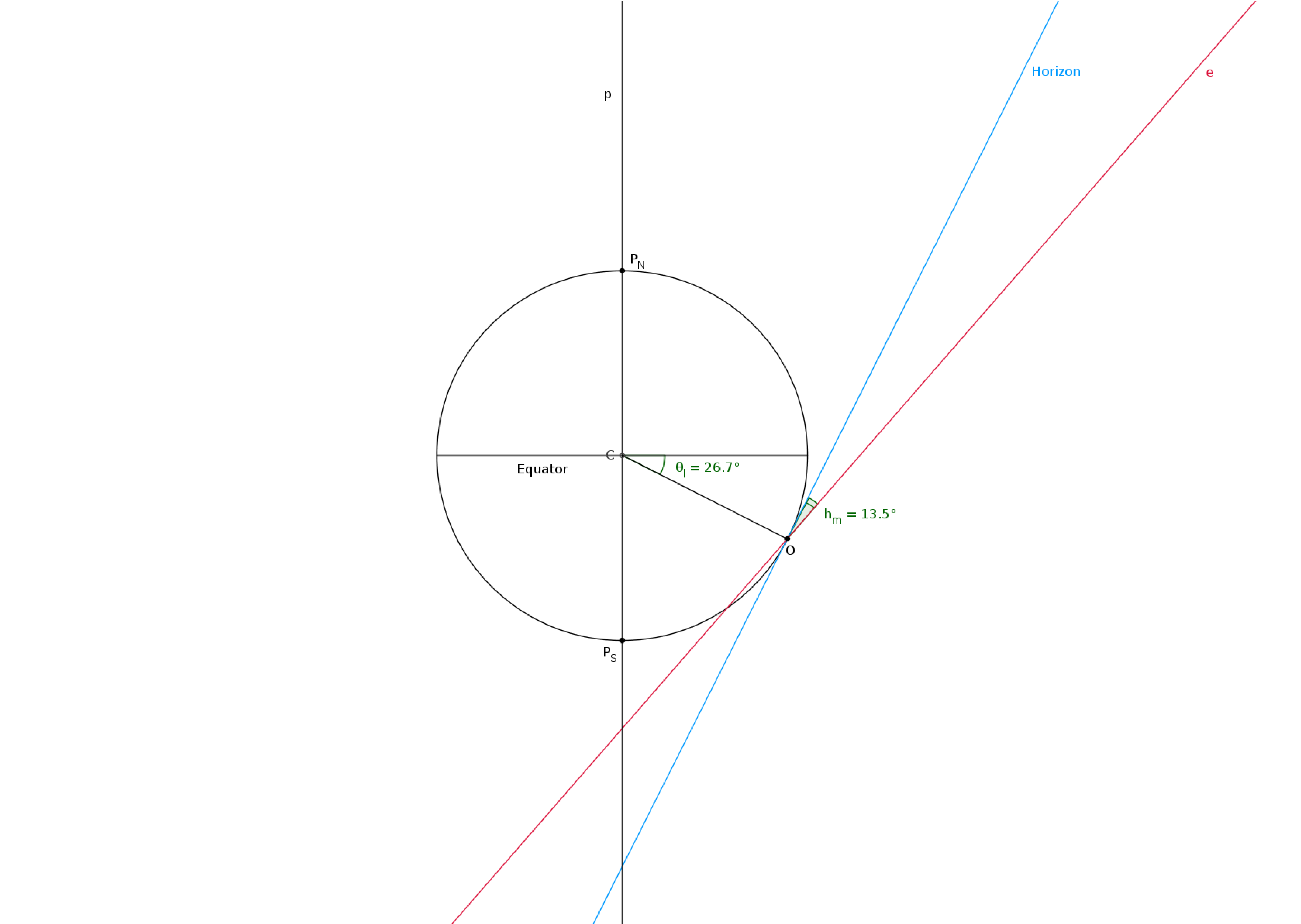}}
\fbox{\includegraphics[width=\columnwidth]{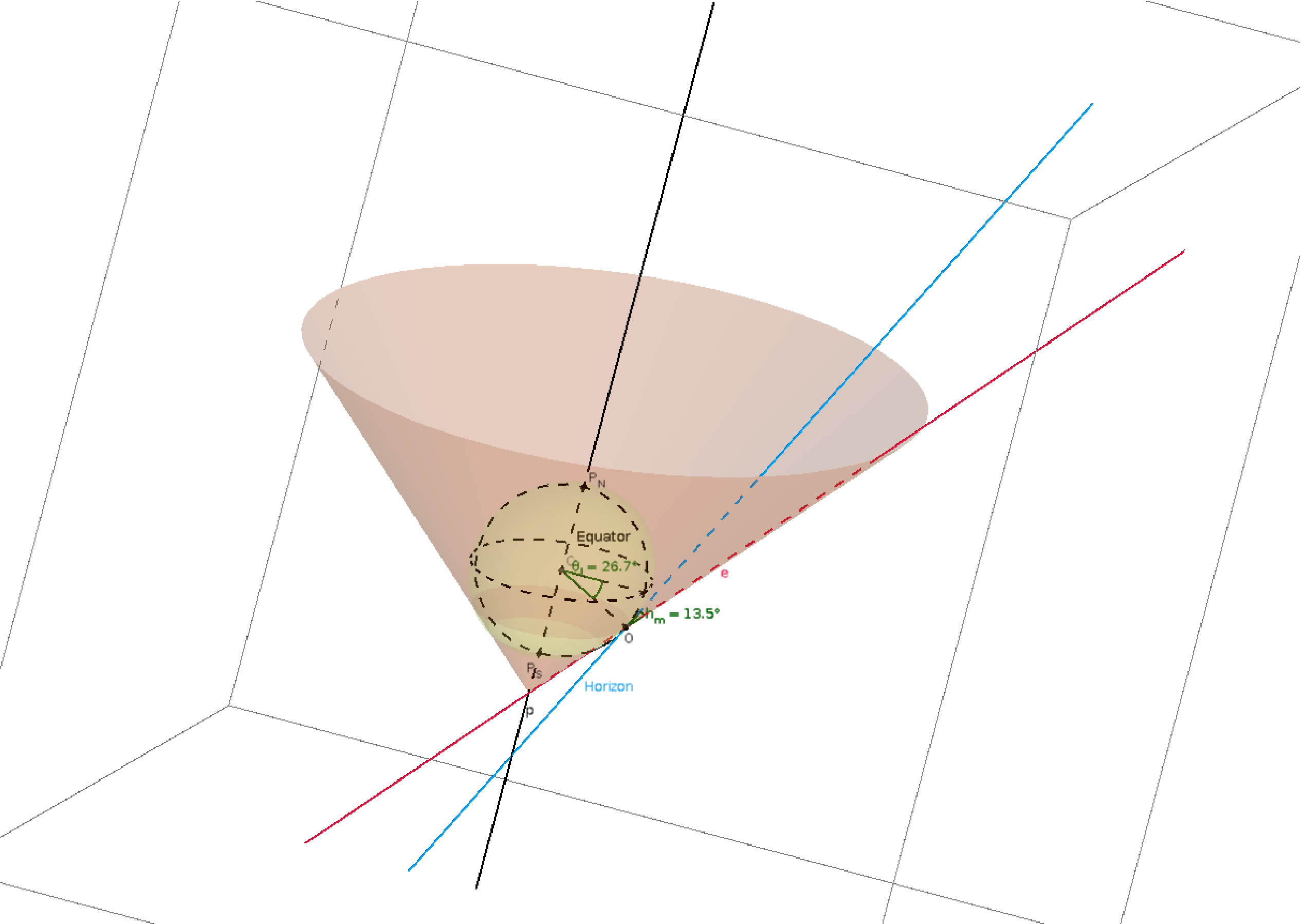}}
\caption{In the left panel, a 2D sketch of the Earth, with the polar axes, the South African (mean) antenna latitude $O$, the horizon line (in blue and the label ``Horizon'') and the lowest angle-limit line (in red end the label ``$e$'') for the antenna. In the right panel, the rotation simulates the Earth rotation and a shadow-cone is traced by the red line.}
 \label{fig: cono d'ombra}
\end{figure*}

Having collected the information about  GRBs detected by satellite missions on  $4 \pi$ sterad, it is necessary to select which part of the sky will be observable by the SKA  from its position on the Earth. Starting from a circle, pole axis and equator can be traced, and it is possible to take into account the (roughly averaged) latitudes where antennas will be placed, i.e., $\sim -30.7^\circ$ in South Africa and $\sim -26.7^\circ$ in Australia, on its edge. The point on the latitude closest to the equator is chosen to draw a tangent (the \textit{horizon line}). Now if the circle is rotated around the pole axis, a sphere and a double cone whose vertex is on the pole axes (below the south pole) is traced out. For our case, only the cone which contains the sphere can be shaded. As a last step, the minimum  elevation of the SKA dish antenna, i.e. $\sim 13.5^\circ$\footnote{The minimum  elevation must be $ < 15^\circ$, as imposed by constraints \citep{Baseline.Design}}, is taken into account. This angle must be added from the horizon line, because of a structural mechanical limit of the dish antennas, so we must consider an \textit{elevation line}. For the sake of  clarity, a radio telescope is generally used to observe at higher elevation (e.g., $\gtrsim 20$) to have a cleaner signal without emission effects due to the atmosphere which decrease the intensity of the observed radiations. These steps are represented  in Figure \ref{fig: cono d'ombra} to show the final shadow-cone for the SKA. 
%
%
\begin{table*}
 
\caption{This table is a collection and merging of all catalogues mentioned in this paper. In the columns, respectively: RA and DEC in degrees; detection year, month and day; name of the sample related to a given satellite; source name. For more details see Appendix A. The full table is available online.} 
 \centering
 \begin{tabular}{c c c c c c r} 
  \hline
  \hline\
  RA  &   DEC & Year &   Month    & Day  & Cat. & \multicolumn{1}{c}{GRB Name}  \\  
  deg &   deg &      &            &      &      &                               \\ 
 \hline
 
  174.680  &    -44.320  &   1990  &   01  &     18.73618  &    GRA    &      GRB900118   \\
   91.160  &    -82.010  &   1990  &   01  &     20.85941  &    GRA    &      GRB900120   \\
  113.280  &     27.890  &   1990  &   01  &     23.07098  &    GRA    &     GRB900123A   \\
  357.190  &    -38.560  &   1990  &   01  &     23.78091  &    GRA    &     GRB900123B   \\
  131.630  &    -38.180  &   1990  &   01  &     26.75309  &    GRA    &      GRB900126   \\
  338.040  &     35.360  &   1990  &   02  &     22.49763  &    GRA    &      GRB900222   \\
  124.580  &     38.840  &   1990  &   03  &      8.39725  &    GRA    &     GRB900308A   \\
  323.250  &    -31.780  &   1990  &   03  &     27.43366  &    GRA    &      GRB900327   \\
  206.920  &    -12.250  &   1990  &   04  &      4.74625  &    GRA    &      GRB900404   \\
   85.110  &     82.030  &   1990  &   04  &     13.42455  &    GRA    &     GRB900413B   \\
   \dots   &      \dots  &   \dots & \dots &   \dots       &   \dots   & \multicolumn{1}{c}{\dots} \\
   
 \hline
 \hline
 
 \end{tabular}

  \label{Table: all GRBs}
   
\end{table*}
The final angle to consider, between the red line and the polar axes, in Figure \ref{fig: cono d'ombra}, is 
\begin{equation} \label{eq: elevation}
 \vartheta_s = \vartheta_l + 13.5^\circ = 40.2^\circ \simeq 0.702 \rm{\hspace{0.1cm} rad}\,,
\end{equation}
where the shadow cone has a solid angle $\Omega_s$ calculated as:
\begin{equation}  
 \Omega_s = \int_0^{2\pi} \ud \phi \int_0^{\vartheta_s}{\ud \vartheta \sin{\vartheta}} \,,
\end{equation}
resulting in
\begin{equation} \label{eq: shadow cone}
\Omega_s = 2 \pi \cdot [-\cos{\vartheta}]^{\vartheta_s}_{0} = 1.484 \rm{ \hspace{0.1cm} sr}\,.
\end{equation}
Even if it is only an approximation, this integral gives a satisfying idea of the observable sky above the SKA. \par
\begin{figure*}
\includegraphics[width=\columnwidth]{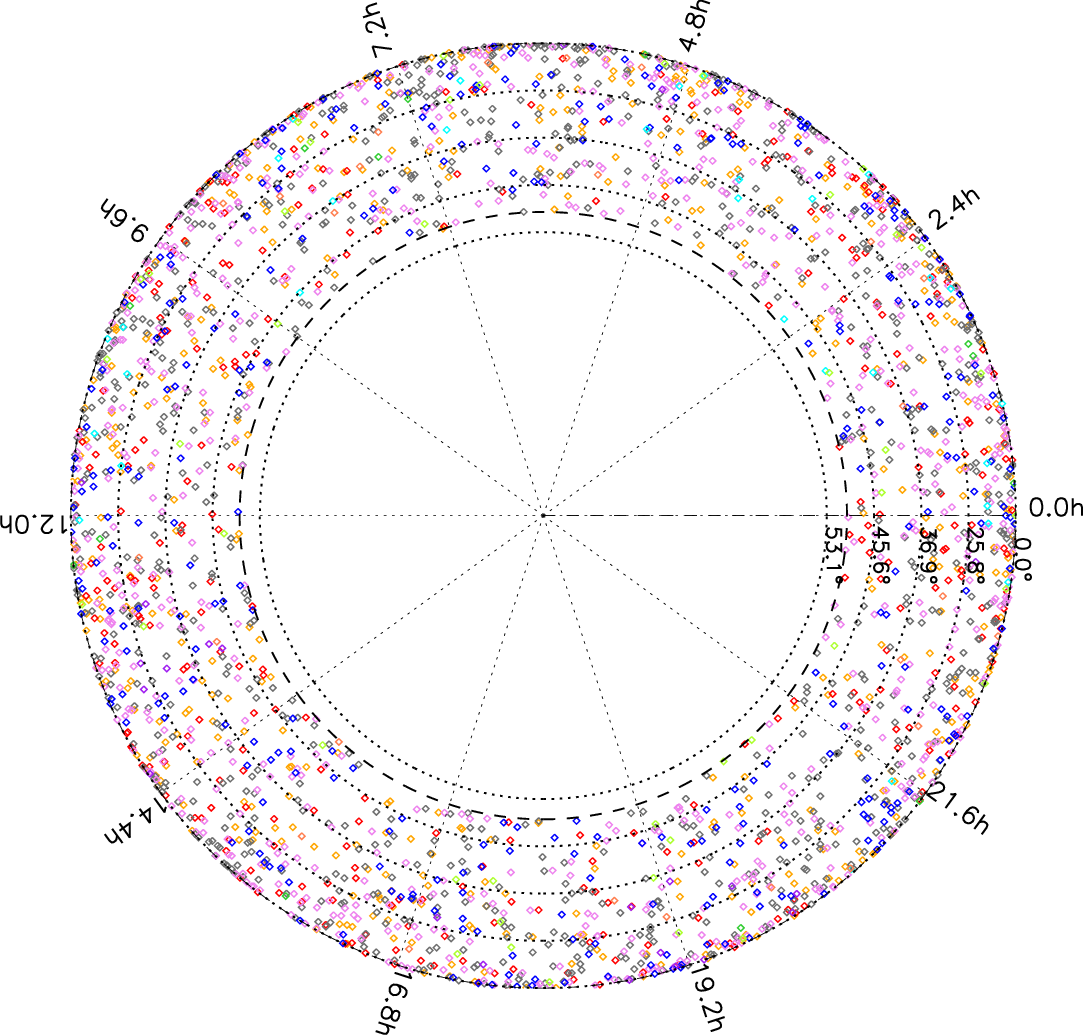}
\includegraphics[width=\columnwidth]{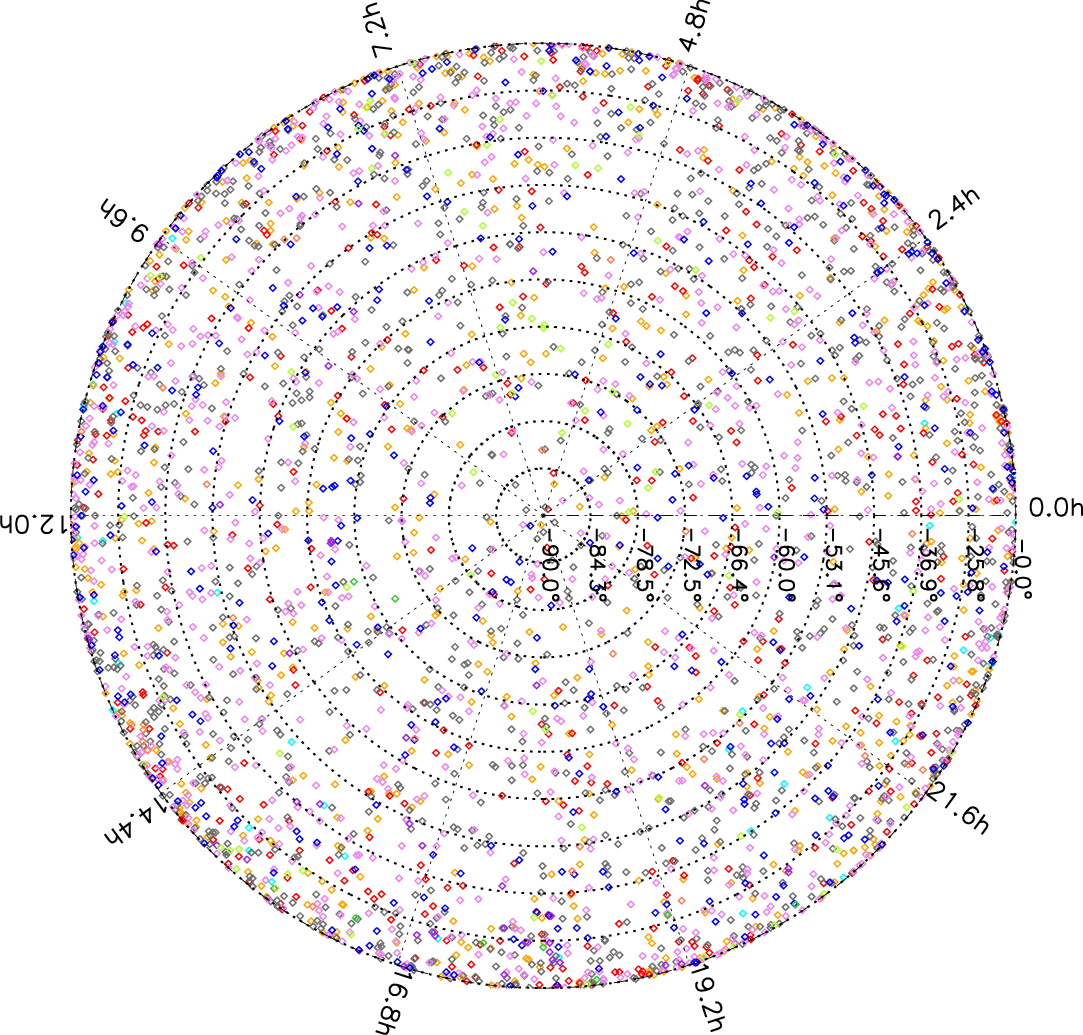}
\caption{Polar plots show 6508 points out of 7516, i.e. GRBs with celestial declination $< 50^\circ$ (dashed circle), observed by GRANAT (coral),  CGRO/BATSE (triggered and non-triggered GRBs are dark grey and violet, respectively), Konus/Wind (yellow-green), BeppoSAX (blue), HETE-2 (cyan), INTEGRAL (purple), \textit{Swift} (red), Agile (lime green) and \textit{Fermi} (orange). The first plot shows the north celestial hemisphere that the SKA can observe, i.e. the ``central hole'' delimited by the black dashed circle is due to the shadow cone, where it will not able to observe; the second plot shows the south celestial hemisphere. RA and DEC are expressed in J2000 coordinates. For the sake of clarity, we have avoided plotting the same sources in different catalogues repeatedly, e.g. Ulysses sources do not appear among these points because they are already plotted together with triggered CGRO/BATSE GRBs.}
 \label{fig: polar plots}
\end{figure*}

 
To conclude this section, Figure \ref{fig: polar plots} shows 6508 GRBs which occurred from 18th January 1990 to 12th May 2014 in two polar plots, corresponding to the north and south celestial hemispheres that the SKA will be able to observe. More details about the developed routine are in Sec. \hyperlink{sec: Appendix A}{Appendix A}.
%
%
%
\begin{table}
\caption{ Values of the angular coefficients of Fig. \ref{fig: TimePlot} are reported.}
 \begin{tabular}{l  c}
  \hline
  \hline
  \multirow{2}{*}{Instrument}  &  Angular    \\
                               & Coefficient \\
  \hline
   GRANAT     &  21.17   \\
   Ulysses    &   66.36  \\
   COMPTEL    &   5.78   \\
   BATSE (t)  &  295.78  \\
   BATSE (nt) &  216.93  \\
   Konus/Wind &  16.01   \\
   BeppoSAX   & 159.28   \\
   HETE-2     &  16.99   \\
   INTEGRAL   &  9.80    \\
   \textit{Swift}      &   90.85  \\
   Agile      &  33.82   \\
   \textit{Fermi}      &   230.97 \\
   
   \hline \hline
   
 \end{tabular}
 
 \label{Tab: angular coefficients}
 
\end{table}
%
\begin{figure*}
\centering
\includegraphics[width=\textwidth]{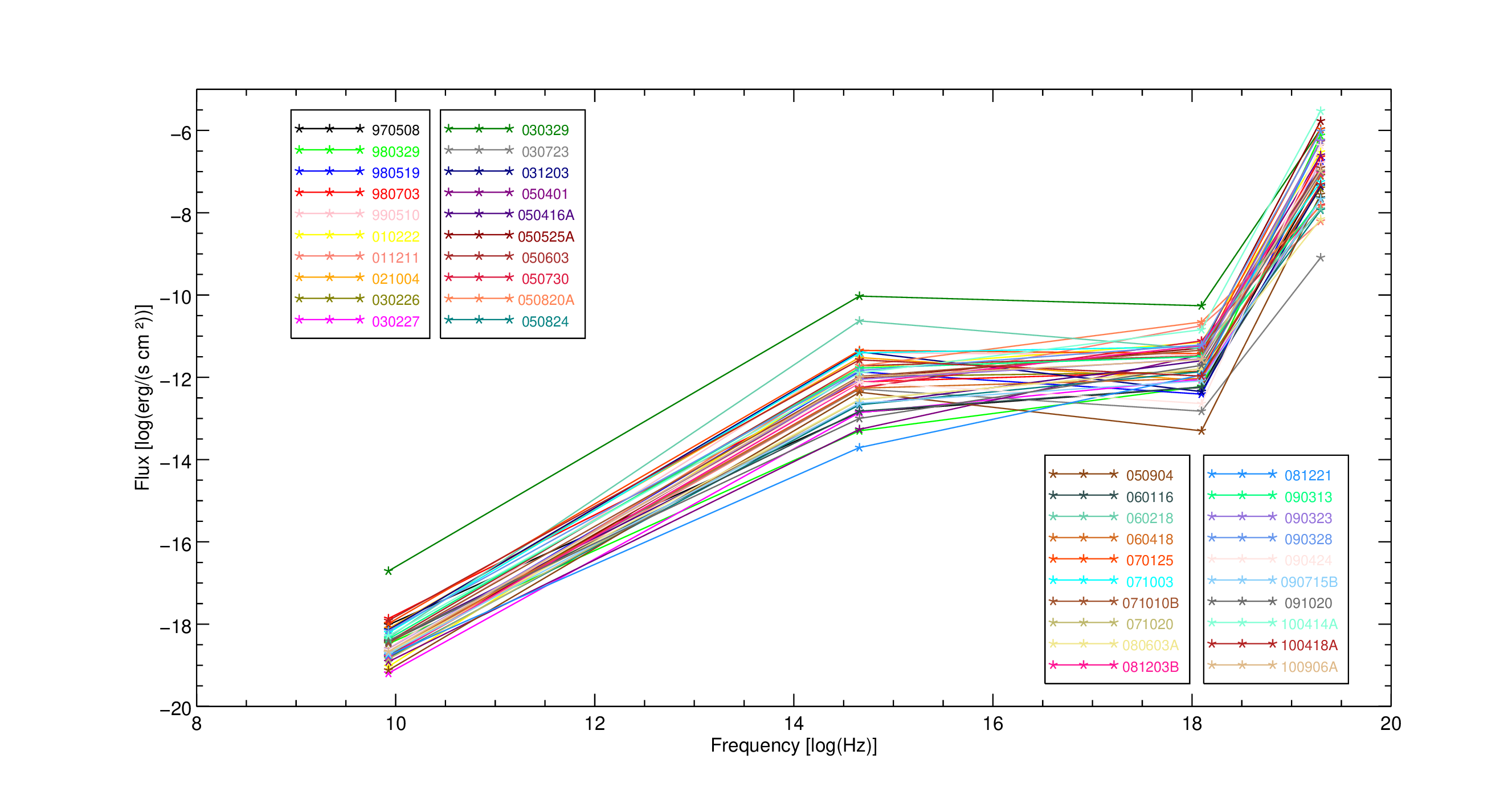}
\caption{Plot of fluxes reported in table \ref{Tab: tutti flussi}. The point are plotted only if the row shows four values at the various frequencies (19.6 EHz, 1.3 EHz, $4.56 \cdot 10^2$ THz, 8.46 GHz).}
 \label{fig: 4 freq}
\end{figure*}


\onecolumn

\begin{landscape}
\begin{longtable}{c c c c c c c c c c c c}
\caption{This table is a fusion between Tables 1 and 4 from \citet{Chandra2012}, but the original $S_{\rm{15-150}}$ fluence, the R-band optical and the 4.86 GHz radio flux densities have been converted into \Flux units, as indicated in Appendix B. The first three columns indicate the GRB name, right ascension and celestial declination. Forth, fifth and sixth columns with Y or N indicate wheter a GRB has a signal in the X, optical and radio band. The last five columns are the fluxes in the 15 - 150 keV and 0.3 - 10 keV energy ranges, at the optical R-band, at the 4.86 GHz radio band, redshift and equivalent isotropic bolometric energy. The apex ``11h'' indicates the value of the flux at 11 hour since the burst. The symbol ``-'' indicates not defined values. } \\

  \hline
  \hline

     Name    &        RA     &      DEC    &    X    &  O    &  R   &    $F_{\rm{15 - 150}}$  &  $F_{\rm{0.3 - 10}}^{11 \rm{h}}$   &   $F_{\rm{R-band}}^{11 \rm{h}}$      &       $F_{\rm{radio}}$   &     $z$    &       $E_{\rm{iso}}^{\rm{bol}}$  \\  
     
\hline

\endfirsthead

\multicolumn{12}{l}{\footnotesize\itshape\tablename~\thetable: ...continues from the previous page} \\

\hline \hline
     Name    &        RA     &      DEC    &    X    &  O    &  R   &    $F_{\rm{15 - 150}}$  &  $F_{\rm{0.3 - 10}}^{11 \rm{h}}$   &   $F_{\rm{R-band}}^{11 \rm{h}}$      &       $F_{\rm{radio}}$   &     $z$    &       $E_{\rm{iso}}^{\rm{bol}}$  \\  

\hline

\endhead

   \hline \hline
   \multicolumn{12}{r}{\footnotesize\itshape\tablename~\thetable:
      continue into the next page...} \\
\endfoot

   \hline \hline
   \multicolumn{12}{r}{\footnotesize\itshape\tablename~\thetable:
      End table} \\
\endlastfoot

   970508   &    06:53:49.2   &     +79:16:19   &  Y   &  Y   &  Y   &     7.92857e-08   &       5.70000e-13   &        1.41304e-13   &        9.58000e-19   &    0.835   &      7.10E+51  \\
   970828   &    18:08:31.7   &     +59:18:50   &  Y   &  N   &  Y   &     7.89116e-07   &       1.99000e-12   &                -   &        1.44000e-19   &    0.958   &      2.96E+53  \\
   980329   &    07:02:38.0   &     +38:50:44   &  Y   &  Y   &  Y   &     7.51724e-07   &       6.00000e-13   &        5.00000e-14   &        3.32000e-19   &     2.95   &      2.10E+54  \\
   980519   &    23:22:21.5   &     +77:15:43   &  Y   &  Y   &  Y   &     1.81000e-07   &       3.90000e-13   &        1.35652e-12   &        2.05000e-19   &    -   &         -  \\
   980703   &    23:59:06.7   &     +08:35:07   &  Y   &  Y   &  Y   &     2.23333e-07   &       1.40000e-12   &        7.71739e-13   &        1.37000e-18   &    0.966   &      6.90E+52  \\
   981226   &    23:29:37.2   &     -23:55:54   &  Y   &  N   &  Y   &     1.34000e-08   &       2.80000e-13   &                -   &        1.37000e-19   &     1.11   &       5.9E+51  \\
   990510   &    13:38:07.1   &     -80:29:48   &  Y   &  Y   &  Y   &     1.61333e-07   &       3.47000e-12   &        3.95435e-12   &        2.55000e-19   &    1.619   &      1.78E+53  \\
   991208   &    16:33:53.5   &     +46:27:21   &  X   &  Y   &  Y   &     1.34833e-06   &               -   &        8.53261e-12   &        1.80400e-18   &    0.706   &      1.10E+53  \\
   000210   &    01:59:15.6   &     -40:39:33   &  Y   &  N   &  Y   &     4.09000e-06   &       3.10000e-13   &                -   &        9.30000e-20   &    0.850   &      2.00E+53  \\
  000301C   &    16:20:18.6   &     +29:26:36   &  X   &  Y   &  Y   &     3.47000e-07   &               -   &        1.44130e-12   &        5.20000e-19   &    2.034   &      4.37E+52  \\
   000418   &    12:25:19.3   &     +20:06:11   &  X   &  Y   &  Y   &     7.16667e-07   &               -   &        6.97826e-13   &        1.08500e-18   &    1.119   &      7.51E+52  \\
   000911   &    02:18:34.4   &     +07:44:29   &  X   &  Y   &  Y   &     1.65000e-08   &               -   &        1.24565e-12   &        2.63000e-19   &    1.059   &      8.80E+53  \\
   000926   &    17:04:09.7   &     +51:47:10   &  X   &  Y   &  Y   &     1.45200e-06   &               -   &        2.29130e-12   &        6.29000e-19   &    2.039   &      2.70E+53  \\
   001007   &    04:05:54.3   &     -21:53:46   &  X   &  Y   &  Y   &     1.45333e-07   &               -   &        3.03261e-12   &        2.22000e-19   &    -   &         -  \\
   001018   &    13:14:10.3   &     +11:48:32   &  X   &  X   &  Y   &     6.38710e-07   &               -   &                -   &        5.90000e-19   &    -   &         -  \\
   010222   &    14:52:12.5   &     +43:01:06   &  Y   &  Y   &  Y   &     2.97059e-07   &       7.05000e-12   &        1.67174e-12   &        9.30000e-20   &    1.477   &      1.33E+54  \\
   010921   &    22:55:59.9   &     +40:55:53   &  X   &  Y   &  Y   &     2.96667e-07   &               -   &        3.12391e-12   &        2.29000e-19   &    0.450   &      9.00E+51  \\
   011030   &    20:43:35.4   &     +77:17:20   &  Y   &  N   &  Y   &            -      &       1.80000e-11   &                -   &        1.39000e-19   &        3   &         -  \\
   011211   &    11:15:18.0   &     -21:56:56   &  Y   &  Y   &  Y   &     6.20000e-09   &       1.80000e-11   &        5.17391e-13   &        1.62000e-19   &    2.140   &      6.30E+52  \\
   020305   &   12:42:27.94   &   -14:18:11.8   &  X   &  Y   &  Y   &     2.65182e-08   &               -   &        1.46957e-12   &        7.60000e-20   &      2.8   &         -  \\
   020405   &   13:58:03.12   &   -31:22:21.9   &  X   &  Y   &  Y   &     1.23750e-06   &               -   &        1.26739e-12   &        1.13000e-19   &    0.690   &      1.10E+53  \\
  020819B   &   23:27:19.47   &   +06:15:56.0   &  X   &  N   &  Y   &     7.94000e-08   &               -   &                -   &        2.91000e-19   &    0.410   &      7.90E+51  \\
   021004   &   00:26:54.68   &   +18:55:41.6   &  Y   &  Y   &  Y   &     2.26000e-08   &       8.40000e-13   &        3.02826e-12   &        7.80000e-19   &    2.330   &      3.80E+52  \\
   021206   &   16:00:46.85   &   -09:42:33.5   &  X   &  N   &  Y   &     1.32000e-05   &               -   &                -   &        1.37000e-19   &    -   &         -  \\
  030115A   &   11:18:32.60   &   +15:02:59.0   &  X   &  Y   &  Y   &     2.62500e-08   &               -   &        1.04348e-13   &        8.30000e-20   &    2.500   &      3.91E+52  \\
   030226   &   11:33:04.93   &   +25:53:55.3   &  Y   &  Y   &  Y   &     3.92754e-08   &       1.35000e-12   &        1.03913e-12   &        1.71000e-19   &    1.986   &      1.20E+53  \\
   030227   &   04:57:33.00   &   +20:29:09.0   &  Y   &  Y   &  Y   &     2.16061e-08   &       9.00000e-13   &        1.39130e-13   &        6.40000e-20   &    -   &         -  \\
   030329   &   10:44:49.96   &   +21:31:17.4   &  Y   &  Y   &  Y   &     1.07619e-06   &       5.48600e-11   &        9.40870e-11   &        1.95670e-17   &    0.169   &      1.80E+52  \\
   030418   &   10:54:33.69   &   -07:01:40.8   &  X   &  Y   &  Y   &     9.72727e-09   &               -   &        5.10870e-13   &        6.90000e-20   &    -   &         -  \\
   030723   &   21:49:24.40   &   -27:42:47.4   &  Y   &  Y   &  Y   &     8.12903e-10   &       1.50000e-13   &        5.19565e-13   &        2.04000e-19   &    -   &         -  \\
   031203   &   08:02:30.36   &   -39:51:00.1   &  Y   &  Y   &  Y   &     4.43333e-08   &       4.50000e-13   &        4.16304e-12   &        7.24000e-19   &    0.105   &      1.15E+50  \\
   050401   &   16:31:28.82   &   +02:11:14.8   &  Y   &  Y   &  Y   &     2.49091e-07   &       3.51000e-12   &        5.43478e-14   &        1.22000e-19   &    2.898   &      3.20E+53  \\
  050416A   &   12:33:54.60   &   +21:03:24.0   &  Y   &  Y   &  Y   &     1.22333e-07   &       2.53000e-12   &        2.15217e-13   &        3.73000e-19   &    0.650   &      1.00E+51  \\
  050509C   &   12:52:53.94   &   -44:50:04.1   &  Y   &  Y   &  Y   &     7.56000e-09   &               -   &        4.08696e-13   &        3.44000e-19   &    -   &         -  \\
  050525A   &   18:32:32.57   &   +26:20:22.5   &  Y   &  Y   &  Y   &     1.70000e-06   &       5.15000e-12   &        9.43478e-13   &        1.64000e-19   &    0.606   &      2.04E+52  \\
   050603   &   02:39:56.89   &   -25:10:54.6   &  Y   &  Y   &  Y   &     5.30000e-07   &       3.30000e-12   &        1.90435e-12   &        3.77000e-19   &    2.821   &      5.00E+53  \\
  050713B   &   20:31:15.50   &   +60:56:38.4   &  Y   &  N   &  Y   &     2.54400e-08   &       1.18000e-11   &                -   &        3.43000e-19   &    -   &         -  \\
   050730   &   14:08:17.13   &   -03:46:16.7   &  Y   &  Y   &  Y   &     1.51592e-08   &       7.67000e-12   &        5.60870e-13   &        2.12000e-19   &    3.968   &      9.00E+52  \\
  050820A   &   22:29:38.11   &   +19:33:37.1   &  Y   &  Y   &  Y   &     1.43333e-08   &       2.21200e-11   &        1.89565e-12   &        1.50000e-19   &    2.615   &      2.00E+53  \\
   050824   &   00:48:56.05   &   +22:36:28.5   &  Y   &  Y   &  Y   &     1.15652e-08   &       1.37000e-12   &        2.15217e-13   &        1.52000e-19   &    0.830   &      1.50E+51  \\
   050904   &   00:54:50.79   &   +14:05:09.4   &  Y   &  Y   &  Y   &     2.77586e-08   &       5.00000e-14   &        4.34783e-13   &        7.60000e-20   &    6.290   &      1.30E+54  \\
   060116   &   05:38:46.28   &   -05:26:13.1   &  Y   &  Y   &  Y   &     2.27358e-08   &       5.80000e-13   &        1.50000e-13   &        3.63000e-19   &    -   &         -  \\
   060218   &   03:21:39.68   &   +16:52:01.8   &  Y   &  Y   &  Y   &     1.22656e-08   &       4.88000e-12   &        2.35261e-11   &        4.71000e-19   &    0.033   &      2.90E+48  \\
   060418   &   15:45:42.40   &   -03:38:22.8   &  Y   &  Y   &  Y   &     8.08738e-08   &       9.40000e-13   &        5.43478e-13   &        2.16000e-19   &    1.490   &      1.00E+53  \\
   070125   &   07:51:17.77   &   +31:09:04.1   &  Y   &  Y   &  Y   &     1.07333e-06   &       3.78000e-12   &        4.51522e-12   &        1.02800e-18   &    1.548   &      9.55E+53  \\
  070612A   &   08:05:29.61   &   +37:16:15.1   &  X   &  Y   &  Y   &     2.87263e-08   &               -   &        7.16957e-12   &        1.02800e-18   &    0.617   &      9.12E+51  \\
   071003   &   20:07:24.22   &   +10:56:50.0   &  Y   &  Y   &  Y   &     5.60811e-08   &       5.57000e-12   &        3.91522e-12   &        6.16000e-19   &    1.604   &      3.24E+53  \\
  071010B   &   10:02:09.26   &   +45:43:50.3   &  Y   &  Y   &  Y   &     1.22222e-07   &       4.56000e-12   &        1.05000e-12   &        3.41000e-19   &    0.947   &      2.60E+52  \\
   071020   &   07:58:39.78   &   +32:51:40.4   &  Y   &  Y   &  Y   &     5.75000e-07   &       1.61000e-12   &        2.84783e-13   &        1.41000e-19   &    2.146   &      8.91E+52  \\
  080603A   &   18:37:37.97   &   +62:44:38.9   &  Y   &  Y   &  Y   &     7.00000e-09   &       1.58000e-12   &        2.89130e-13   &        2.07000e-19   &    1.687   &         -  \\
  081203B   &   15:15:11.67   &   +44:25:42.9   &  Y   &  Y   &  Y   &     9.13044e-08   &       5.87000e-12   &        7.32609e-13   &        1.62000e-19   &    -   &         -  \\
   081221   &   01:03:10.20   &   -24:32:53.2   &  Y   &  Y   &  Y   &     5.55882e-07   &       1.14000e-12   &        1.95652e-14   &        1.74000e-19   &    -   &         -  \\
   090313   &   13:13:36.21   &   +08:05:49.8   &  Y   &  Y   &  Y   &     2.12676e-08   &       3.19000e-12   &        1.67174e-12   &        4.35000e-19   &    3.375   &      4.57E+52  \\
   090323   &   12:42:50.29   &   +17:03:11.6   &  Y   &  Y   &  Y   &     5.50376e-07   &       2.81000e-12   &        8.95652e-13   &        2.43000e-19   &     3.57   &      4.10E+54  \\
   090328   &    06:02:39.6   &   -41:53:03.2   &  Y   &  Y   &  Y   &     9.47368e-07   &       6.14000e-12   &        1.46522e-12   &        6.86000e-19   &    0.736   &      1.00E+53  \\
   090423   &   09:55:33.29   &   +18:08:57.8   &  Y   &  Y   &  Y   &     6.25000e-08   &       5.00000e-13   &                -   &        5.00000e-20   &    8.260   &      1.10E+53  \\
   090424   &   12:38:05.11   &   +16:50:15.1   &  Y   &  Y   &  Y   &     4.36000e-07   &       2.30000e-13   &        7.65217e-13   &        2.36000e-19   &    0.544   &      4.47E+52  \\
  090715B   &   16:45:21.53   &   +44:50:20.0   &  Y   &  Y   &  Y   &     2.13962e-08   &       8.20000e-13   &        2.34783e-13   &        1.91000e-19   &    3.000   &      2.36E+53  \\
  090902B   &    17:39:45.6   &   +27:19:26.6   &  Y   &  Y   &  Y   &         -         &       4.70000e-12   &        3.19565e-13   &        8.40000e-20   &    1.883   &      3.09E+54  \\
   091020   &   11:42:55.21   &   +50:58:42.2   &  Y   &  Y   &  Y   &     9.74359e-08   &       1.95000e-12   &        1.00000e-13   &        3.99000e-19   &    1.710   &      4.56E+52  \\
  100413A   &   17:44:53.22   &   +15:50:02.4   &  Y   & Y?   &  Y   &     3.24607e-08   &       1.26000e-12   &                -   &        8.00000e-20   &      3.5   &         -  \\
  100414A   &   12:48:29.96   &   +08:41:34.9   &  Y   &  Y   &  Y   &     2.97692e-06   &       1.43700e-11   &        1.27174e-12   &        5.24000e-19   &    1.368   &      7.79E+53  \\
  100418A   &   17:05:27.18   &   +11:27:40.1   &  Y   &  Y   &  Y   &     4.85714e-08   &       1.07000e-12   &        2.65435e-12   &        1.21800e-18   &    0.620   &      5.20E+50  \\
  100906A   &   01:54:44.15   &   +55:37:50.5   &  Y   &  Y   &  Y   &     1.05263e-07   &       2.71000e-12   &        1.01304e-12   &        2.15000e-19   &    1.727   &      1.34E+53

 \label{Tab: tutti flussi}                                                     
 \end{longtable}
\end{landscape}

\twocolumn


\subsection{The  Spectral Energy Distribution from radio to gamma band}  \label{Sec: A suggestion for a Spectral Energy Distribution from radio to gamma band}
In order to fix ideas for GRB detections in radio band, we need to relate GRB gamma emission to GRB radio emission in a reasonable way. To do this, we take into account the Tables 1 and 4 in \cite{Chandra2012}. By converting all detected (or extrapolated) data from fluence and flux densities into fluxes [\Flux], we obtain Table \ref{Tab: tutti flussi} which gathers four frequencies at 19.6 EHz, 1.3 EHz, $4.56 \cdot 10^2$ THz, 8.46 GHz, corresponding to the ranges 15\,-\,150 keV, 0.3\,-\,10 keV, the optical R-band and the obvious radio frequency. 
The Table shows 64 GRBs and their corresponding celestial coordinates observed at different frequencies. The adopted  procedure to obtain  fluxes is better described in \hyperlink{sec: Appendix B}{Appendix B}, where the time considered for conversion from fluence has been the $T_{90}$. To convert flux densities for radio and optical bands, their acquisition bandwidths have been considered. The symbols ``Y'' and ``N'' in columns 3$^{th}$, 4$^{th}$ and 5$^{th}$ indicate if the signal is detected at X-ray, optical or radio frequencies. \\
Figure \ref{fig: 4 freq} shows a plot with the four frequencies, i.e. fluxes vs 19.6 EHz, 1.3 EHz, $4.56 \cdot 10^2$ THz, 8.46 GHz. Data are taken from Table \ref{Tab: tutti flussi}, but only where fluxes have all four values. At the end, the log\,-\,log plot reports 40 SEDs with three slopes and it is easy to see that the flux decreases with the wavelength. Implementing a linear fit for each gap and each source, we obtain the means between the observable frequencies. The angular coefficients (or spectral indexes) are 3.89461 between the gamma and X bands, 0.122765 between X-ray and R bands and, finally, 1.34954 between R-band and 8.46 GHz. \par
\begin{figure*}
\centering
\includegraphics[width=\textwidth]{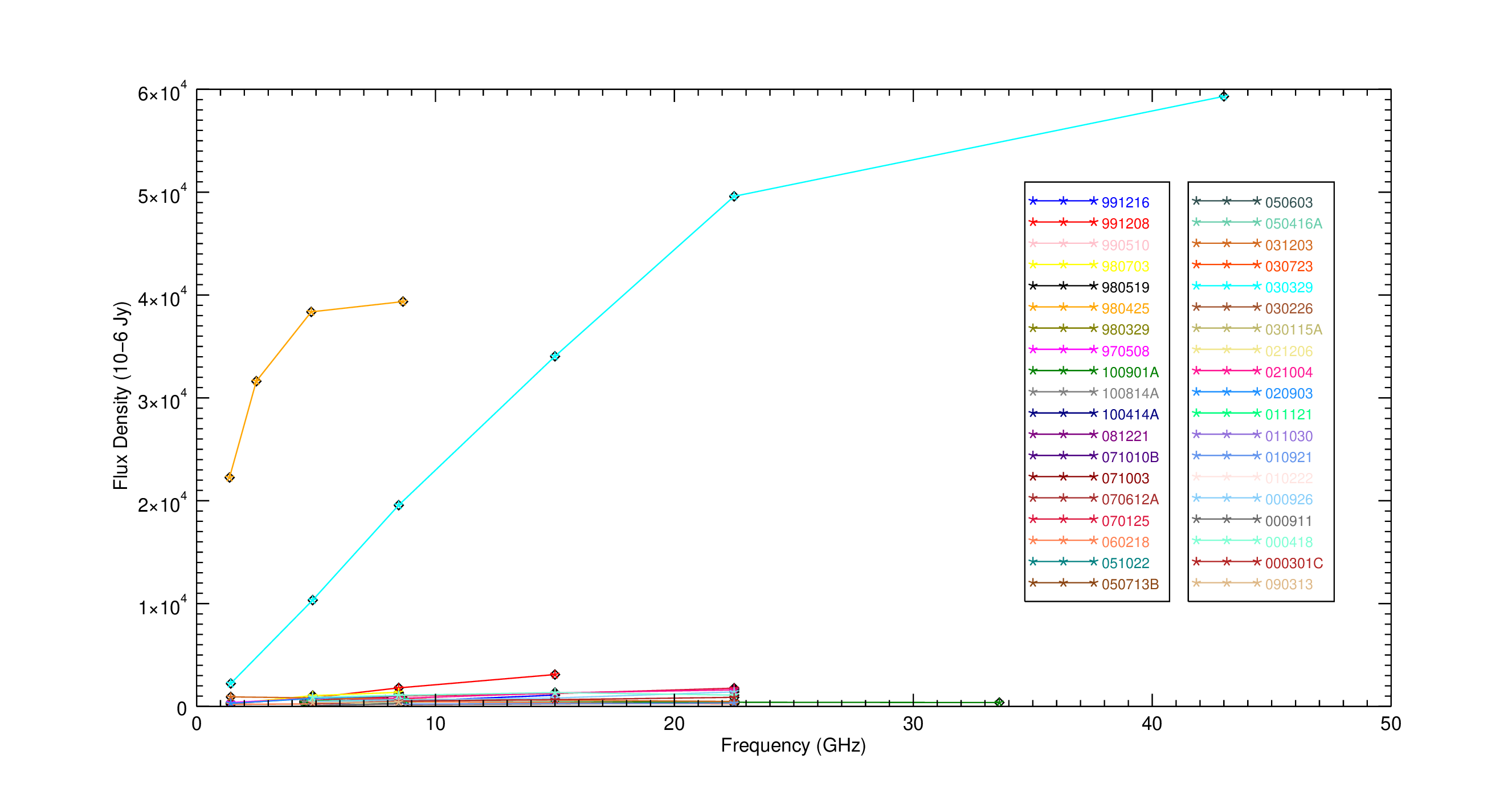}
\includegraphics[width=\textwidth]{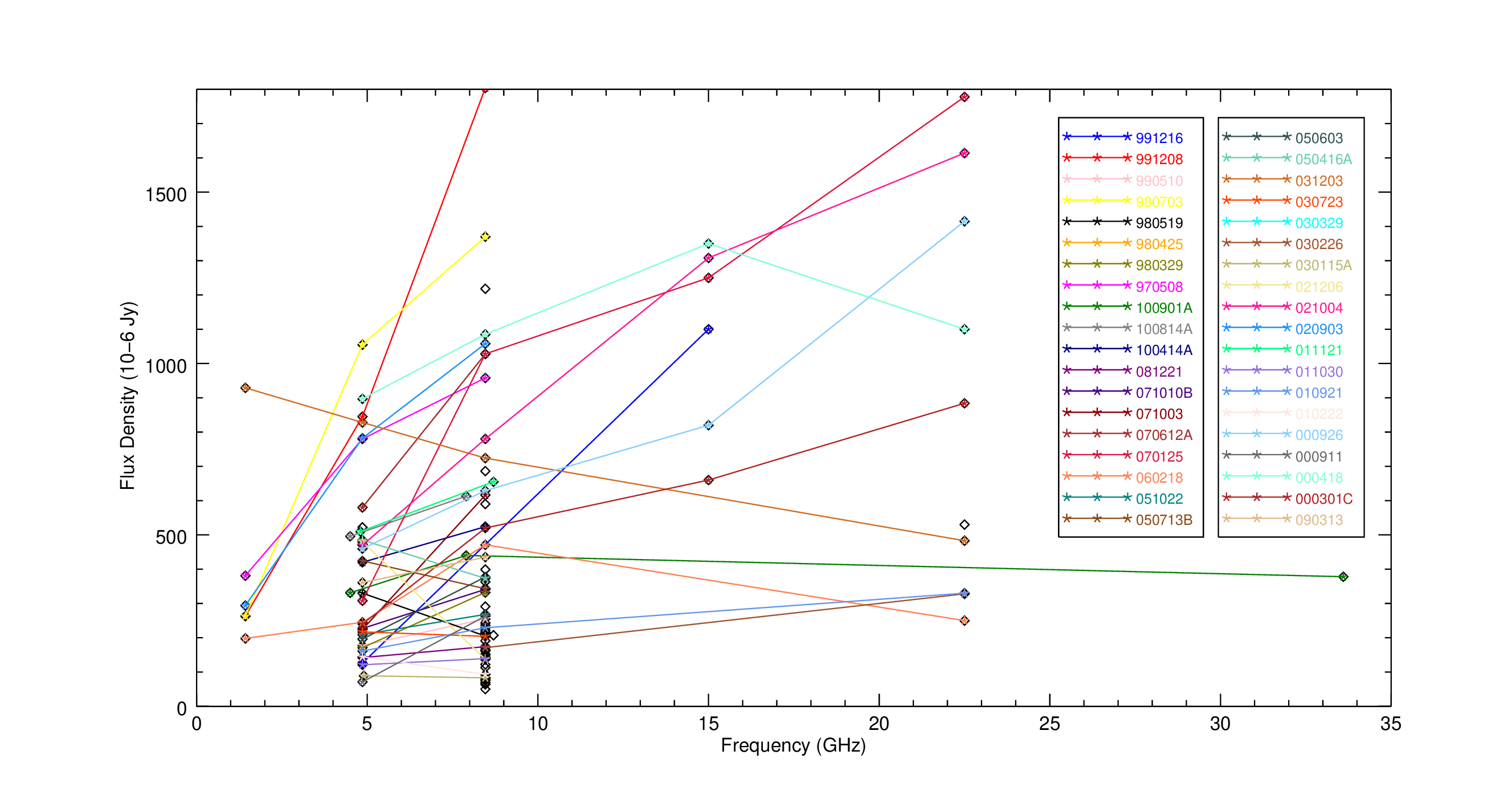}
\caption{These plots  show the fluxes densities vs frequency of GRBs reported in Table 4 by  \citet{Chandra2012}. The lower panel is simply a zoom of the upper panel. Each GRB is fitted by a linear fit. The isolated points are GRB observed only at one frequency. In the internal picture frames,  the GRB names are reported and  associated with different colors. }
 \label{fig: radio freq}
\end{figure*}

Limited to Table 4 (always in \cite{Chandra2012}), we have carried out a method similar to previous one, collecting different flux densities of the same GRB observed within the radio band 1.4\,-\,43 GHz, and plotting their data vs frequency. The points are shown in Figure \ref{fig: radio freq}.  If a GRB was detected at least twice, we have been able to trace a line to connect points each other. By making a linear fit for each set, we obtain some means in this range: 124.87, 55,60 and 36.50. The differences among these values depend on the considered fits. The first mean concerns all the 38 GRBs in the plot, for the second one we have excluded angular coefficient too steep, negative and irregular trend, in the third we have ignored the too steep slope only. The value 55.60 is a mean among 27 fits and is the most trustworthy, so it will be used in the following. 
%
%
\subsection{The SKA sensitivity and the radio GRB-detection probability}  \label{Sec: The SKA sensitivity and the radio GRB-detection probability}
Considerations about flux extrapolations have been done in Sec. \ref{Sec: A suggestion for a Spectral Energy Distribution from radio to gamma band}, hence we can discuss here about the minimum  flux densities detectable by the SKA. As mentioned, because the sensitivity at 8.46 GHz for this radio telescope will be about $1.56 \cdot 10^{-1}$ $\mu$Jy, we multiply by 3 to have the $3 \sigma$. Thanks to the graphic in Figure \ref{fig: 4 freq} and its mean slopes calculated, we can extrapolate the minimum detectable gamma flux which the SKA will be able to detect, due to the luminosity decay. \\
First of all, we must convert the SKA sensitivity from flux density into flux. To this goal,   we write 
\begin{equation}
\begin{array}{l}
 F_{3 \sigma} = 4.68 \cdot 10^{-1} \, \times \, 9.2 \cdot 10^8 \, \rm{Hz} \, \times 10^{-29} \, \frac{ \rm{erg}} {\rm{cm}^{2} \, \rm{s} \,\, \rm{Hz}} \, =  \\
 \\
 = 4.29 \cdot 10^{-20} \, \rm{erg} \, \rm{cm}^{-2} \, \rm{s}^{-1} \,, 
 \end{array}
 \label{eq: 3 sigma 8.46 GHz}
\end{equation}
where 9200 MHz is the bandpass at this specific frequency and $10^{-29}$ \Flux \, Hz$^{-1}$ is a conversion factor to convert Jy into \Flux \, Hz$^{-1}$. Considering the $\log_{10}$ of  the result in Eq. \ref{eq: 3 sigma 8.46 GHz} and assuming the spectral indexes in Figure \ref{fig: 4 freq}, we obtain an estimate of the minimum flux in the gamma band. This value is $\sim 1.25 \cdot 10^{-8}$ \Flux and corresponds to the minimum flux that a GRB should emit, at 15\,-\,150 keV, in order to be detected at 8.46 GHz by the SKA, that is when the luminosity is decaying and passing through  lower frequencies from the gamma band. \par
Regarding the SKA-Low, we have already mentioned that the planned sensitivity is 2.06 $\mu$Jy hr$^{-1/2}$, so the $3\sigma$ will be 3.57 $\mu$Jy with an observation of 3 hours. By using the previous spectral index found in radio range, 55.60, we can extrapolate an estimated value at 8.46 GHz starting from 150 MHz (which is the central frequency in the observational range of the LFAA). In this way, we extrapolate the minimum flux density of 465.60 $\mu$Jy, corresponding to a flux equal to $1.40 \cdot 10^{-18}$ \Flux  (by using a bandpass at  300 MHz). Therefore, with the method previously used, starting from 8.46 GHz, a GRB might be detected by the LFAA if the flux at 15\,-\,150 keV emits at least $4.08 \cdot 10^{-7}$ \Flux. \\
Since satellite measures are in terms of fluence, we have to divide  the  extrapolated fluxes by a mean $T_{90} \sim 70$ s, to have more common measure units. So that if a GRB is detected by the SKA antennas, it should emit a fluence of $S_{\rm{m, \, mid}} \gtrsim 8.75 \cdot 10^{-7} $ \Fluence for the SKA-Mid and $S_{\rm{m, \, low}} \gtrsim 2.86 \cdot 10^{-5}$ \Fluence for the SKA-Low. \par
In \cite{Chandra2012}, the authors state that, in their observed sample, the ratio between radio detections and non-detections is
\begin{equation} 
 \frac{95}{304} = 0.313\% \,.
\end{equation} 
This ratio is useful for  some considerations. In can be  assumed as   a projection of an effective  distribution for the radio detectability. Despite of this assumption,   Chandra \& Frail  point out  that this ratio can be faulty because of a insufficient sensitivity of current radio telescopes. As a consequence, authors by \cite{Hancock2013} suppose that some radio emissions can be among  radio non-detections.  According to this consideration,  we can apply their results to our calculations, using also our estimated fluence limits. Thus, considering the $S_\gamma$ value for the RF population reported in Table 3 of \cite{Hancock2013}, it is possible to draw GRBs with $S_\gamma \geq 1.6 \cdot 10^{-6}$ \Fluence \, from \textit{Swift} Table considered in this paper. This fluence value has the same magnitude found in \cite{Chandra2012} as a rough threshold between radio detection and non-detections and it is higher than our extrapolated limit. \par 
By limiting the calculation to the \Swift observations taken from the \href{http://swift.gsfc.nasa.gov/archive/grb\_table}{``Swift GRB Table and Lookup''} (time-frame 17$^{\rm{th}}$ December 2004 - 18$^{\rm{th}}$ March 2015) some estimations can be carried out. In this smaller group, there are 951 GRBs in total, where 146 have a declination greater than $50^\circ$ (namely out of the SKA view). The remaining 805 GRBs have $\delta < 50^\circ$, while only 37 of them are without a fluence measured between 15-150 keV (some of them are ``to be confirmed'' (TBC)). It is worth noticing that the percentage of GRBs detected by the \Swift within the sky above the SKA is 
\begin{equation}
 \frac{805}{951} \simeq 85\% \,
\end{equation}
which is close to the percentage found with our larger sample
\begin{equation} \label{eq: percentuale grande}
 \frac{6508}{7516} = 86.6\% \,.
\end{equation} 
Hence,  the restricted group in \Swift catalogue is enough representative for our next discussion. \\
Considering the threshold gamma fluence over which the radio detections found by Chandra \& Frail were about 86\%, i.e. $S_{\rm{15-150}} = 10^{-6}$ \Fluence, it is sraightforward to see that 578 GRBs are above   this limit and have $\delta < 50^\circ$. Therefore,  the percentage of detecting GRBs for the SKA, in radio band and in the ``\Swift world'', is at least 
\begin{equation} \label{eq: probabilita Chandra}
 P_{\rm{CF}} = \frac{578}{805} = 71.8\% \,.
\end{equation}
On the other hand, it is possible to take into account the values calculated by \cite{Hancock2013}, by using again the same catalogue of the \Swift. As mentioned, the median value of fluence at 15-150 keV, associated to their RF GRBs,  is $1.6 \cdot 10^{-6}$ \Fluence. Thus, directly considering the level of the fainter emitters (obviously RB GRBs have higher fluences), the detection rate for the SKA becomes 
\begin{equation} \label{eq: probabilita Hancock}
 P_{\rm{HGM}}\frac{523}{805} = 65.0\% \,.
\end{equation}
By using now the minimum fluence level calculated for the SKA-Mid, we can slightly increase the previous estimations. The GRBs with fluence $\geq 0.88 \cdot 10^{-6}$ \Fluence are 705, hence the rate to detect a GRB radio emitter within the sky above the SKA can be
\begin{equation}  \label{eq: probabilita Mid min}
 P_{\rm{HF, \, min}} = \frac{705}{805} = 87.6\%\,.
\end{equation}
As concerning the LFAA antennas, we can initially consider GRBs detected by the \Swift at any declination, with a fluence $\gtrsim 28.6 \cdot 10^{-6}$ \Fluence, in the 15\,-\,150 keV. In this case, 48 GRBs out of 951 (in this considered new list, 47 GRBs have not fluence measured, or TBC, but probably with a fluence less than $28.6 \cdot 10^{-6}$ \Fluence) are over this limit, thus the detection probability results
\begin{equation}  \label{eq: probabilita Low min}
 P_{\rm{LF, \, min}} = \frac{48}{951} = 5.05\% \,.
\end{equation}
Basically, the estimations of radio detectability of GRB afterglows are greater than 60\% for the SKA-Mid, and $\sim 5\%$ for the SKA1-Low. This last number was expected, because the sensitivity at low frequency is one magnitude less than the higher one \citep{Baseline.Design}. To conclude, we remember that the upper limit calculated in \cite{Hancock2013} for RF GRBs is $\sim 40$ $\mu$Jy at 8.46 GHz, but we have calculated that the planned sensitivity of the SKA-Mid will reach magnitude about nJy. This means that  RF GRBs will not be a unsurmountable obstacle for this new telescope. Only a systematic radio observational  study could point out how much the effective radio-emission percentage can be directly related to GRBs, however the already  achieved theoretical results and simulations  are promising.

\subsection{Serendipitous detection rates for the SKA} \label{sec: Serendipitous detection rate probability for the SKA}
In Sec. \ref{Sec: The SKA sensitivity and the radio GRB-detection probability}, we have reported the estimated probability to detect GRBs at 8.46 GHz and 150 MHz for the SKA. In this section, starting from  our GRB final list, we want to discuss  the detecting rate of  GRBs in a serendipitous way within the planned FoV of the SKA.  However, we are referring to  a potential GRB survey. \\
Let us  take into account the times between the first and the last GRB detections for each instrument, without considering the whole life time of the instrument itself. We are aware that for a more refined calculation,  one should also  take into account the fraction of time in which the instruments are effectively  operating. However  we have not adopted this time interval, so that we can find only a minimum value. In other words, by using this longer time with respect to the shorter operational observation time, focused on GRB monitoring, we can achieve a more conservative estimation and provide  GRB detection rates in  worse cases. 
Here we wants only  to highlight the chance to observe and to study GRB in radio band as a realistic opportunity. \par
In Sec. \ref{Collection of GRB catalogues}, we have discussed how our comprehensive  GRB list has been acquired. We now consider sources with declination lesser than $50^\circ$; these are 6508. Hence, let be $N_{\mathrm{GRB}} = 6508$  the number of GRBs within a spherical cap  equal to $(4 \pi - \Omega_s)$ sterad. As already mentioned, the SKA FoV is $\sim 1$ square degree ($\Omega_{\mathrm{HF}} \simeq 3.05 \cdot 10^{-4}$ sr) at the high-frequency (1 - 20 GHz). Then, we assume a uniform distribution for GRBs detected within the telescope beam and with a time-frame of 24.33 years. This number must be multiplied  by the estimated percentage found in Eq. \ref{eq: probabilita Hancock}, which is the lowest value (hence the worst case) among the equations in Sec. \ref{Sec: The SKA sensitivity and the radio GRB-detection probability}, to detect radio emission from GRBs. Following those steps, we obtain a first detection rate of observing GRBs by the SKA at high frequency: 
\begin{equation}
 R_{\mathrm{HF}} = \frac{N_{\mathrm{GRB}}}{4 \pi - \Omega_s} \cdot \frac{\Omega_{\mathrm{HF}}}{24.33 \hspace{0.1 cm} {\rm yr}} \cdot 0.650 = 4.78 \cdot 10^{-3} \frac{\rm{GRB}}{\rm{yr}}\,. 
\end{equation}
This number shows that the SKA in its high-frequency range has an extremely low probability to serendipitously detect radio transients associated to GRBs. This value can be estimated using all data  considered here  in Table \ref{Table: all GRBs}  in a time-frame of almost 25 years. \par
Let us now carry out the same calculation, but now focusing  on the \textit{Fermi} and \textit{Swift} missions. In our final catalogue, the first \textit{Swift} detection was on 2004/12/17 and the last one is on 2014/05/12, counting 869 GRBs. On the other hand, \textit{Fermi} observed its first GRB on the 2008/07/14 and the last one on the 2014/05/12, having 1359 sources in total. Excluding the sources with a declination $\geq 50^\circ$, they become 1194 by \textit{Fermi} ($N_\mathrm{Fer}$) for 2129 days, and 736 by \textit{Swift} ($N_{\mathrm{Swi}}$) for 3433 days. 
For these two missions, in their own time-frames, the probabilities to detect in a serendipitous way a GRB per year within the high-frequency SKA are:
\begin{eqnarray}
 R_{\mathrm{HF,Fer}} &=& \frac{N_\mathrm{Fer}}{4 \pi - \Omega_s} \cdot \frac{\Omega_{\mathrm{HF}}} {2129/365.25 \rm{\hspace{0.1cm} yr}} \cdot 0.650 = \nonumber \\
                     && \nonumber \\
                     &=& \frac{1194}{11.082 \rm{ \hspace{0.1cm} sr}} \cdot 5.22 \cdot 10^{-5} \hspace{0.1cm} \frac{\mathrm{sr}}{\mathrm{yr}} \cdot 0.650 = \\
                     && \nonumber \\
                     &=& 3.66 \cdot 10^{-3} \frac{\rm{GRB}}{\rm{yr}} \nonumber \,;
\end{eqnarray}
\begin{eqnarray}
R_{\mathrm{HF, Swi}} & = & \frac{N_\mathrm{Swi}}{4 \pi - \Omega_s} \cdot \frac{\Omega_{\mathrm{HF}}}{3433/365.25 \rm{\hspace{0.1cm} yr}} \cdot 0.650 =   \nonumber \\
                 && \nonumber \\
                 &=& \frac{736}{11.082 \rm{ \hspace{0.1cm} sr}} \cdot 3.24 \cdot 10^{-5} \hspace{0.1cm} \frac{\mathrm{sr}}{\mathrm{yr}} \cdot 0.650 = \\ 
                 && \nonumber \\
                 &=& 1.40 \cdot 10^{-3} \frac{\rm{GRB}}{\rm{yr}} \nonumber \,.
\end{eqnarray}
These numbers are still very low. However, looking at Table \ref{Tab: angular coefficients}, even if we use the CGRO mission, that is the mission with the most efficient detection rate, considering both triggered and non-triggered data (4036 GRBs within 3324 days), the result is $ 7.93 \cdot 10^{-3}$ GRB/yr. If we take into account the Earth rotation during an observation in order to increase these numbers, the previous results will be multiplied by a factor 2.5. Therefore, in the best case, the SKA could detect $\sim 2.5 \cdot 7.93 \cdot 10^{-3}$ GRBs per year, at its high frequency. \par 
The situation changes for low-frequency antennas and their FoV. In fact, the SKA-Low has a FoV equal to 200 square degrees, that corresponds to a solid angle $\Omega_{\mathrm{LF}} \simeq 6.1 \cdot 10^{-2}$ sterad. At the low frequency, a slightly smaller spherical cap than before must be consider, because dipole antenna can observe with a scan angle of $\pm 45^\circ$ with respect to the zenith (as constraint reported by \cite{SKA-AA}). Following Eqs. \ref{eq: elevation} and \ref{eq: shadow cone}, we have to add $45^\circ$ to $\vartheta_l$ instead of $13.5^\circ$, in order to obtain the new $\Omega_{s, \mathrm{LF}} \simeq 4.310$ sterad. Thus, with a declination less than to $18^\circ$ ($ \simeq 90^\circ - \vartheta_l  - 45^\circ$), the number of GRBs detected by \textit{Fermi} is 880. Considering also the probability calculated in Eq. \ref{eq: probabilita Low min}, we obtain:
\begin{eqnarray}
 R_{\mathrm{LF,Fer}} &=& \frac{880}{4 \pi - \Omega_{s, \mathrm{LF}}} \cdot \frac{\Omega_{\mathrm{LF}}}{2129/365.25 \rm{\hspace{0.1cm} y}}  \cdot 0.051 =  \nonumber \\
                     && \nonumber \\
                     &=& \frac{880}{8.256 \rm{ \hspace{0.1cm} sr}} \cdot 1.05 \cdot 10^{-2} \hspace{0.1cm} \frac{\mathrm{sr}}{\mathrm{y}} \cdot 0.051 = \\
                     && \nonumber \\
                     &=& 5.71 \cdot 10^{-2} \frac{\rm{GRB}}{\rm{y}} \nonumber \,.
\end{eqnarray}
As previously done for the dishes, we can consider the Earth rotation, thus doubling the SKA-Low FoV. Doing this, we obtain a probability of $1.14 \cdot 10^{-1}$ GRBs per year. \\
Contrary to the SKA dishes, the LFAA could have a greater opportunity to be a GRB-radio monitoring. However,  we have performed a simulation implying a linear extrapolation into the 50-350 MHz range starting from the GHz radio band. Certainly, more self-consistent studies on  GRB low-radio-band observations are needed to draw conclusions on the effective feasibility of GRB radio surveys by the SKA. \par
To complete the subsection, we want to add a more optimistic calculation. Before, we have used  only satellite detections, but if the SKA is used as a  radio-GRB monitor, we should consider all GRBs which lights up towards the Earth, without thinking about satellites. In \cite{Ghirlanda2014rivista}, the authors perform  a simulation to estimate the number of GRB orphan afterglows in the Universe. From this number,  it is possible to estimate the number of all GRBs per year which points towards the Earth, that is $\sim 811.5$. Continuing to assume that the detectable percentage used until now is uniform, we have 
\begin{equation}
 R_{\rm{radio \, GRB}} \sim 811.5 \cdot 0.65 \, \frac{\rm{GRB}}{\rm{yr}} = 527.5 \, \frac{\rm{GRB}}{\rm{yr}} \,.
\end{equation}
With this number, remembering the percentage of GRBs and the spherical cup above the SKA for the SKA-Mid, the previous detection rate becomes
\begin{eqnarray} \label{eq: high freq all rate}
 R_{{\rm{HF,\, radio \, GRB}}} &=& 527.5 \cdot 0.866 \cdot \frac{\Omega_{\rm{HF}}}{4 \pi - \Omega_s} \, \frac{\rm{GRB}}{\rm{yr}} = \nonumber \\
 && \\
 &=& 1.26 \cdot 10^{-2} \, \frac{\rm{GRB}}{\rm{yr}} \nonumber \,.
\end{eqnarray}
As for the detection rate for the SKA1-Low, the probability to have a detectable radio GRB at 150 MHz per year is
\begin{equation}
 R_{\rm{radio \, GRB}} \sim 811.5 \cdot 0.051 \, \frac{\rm{GRB}}{\rm{yr}} = 41.4 \, \frac{\rm{GRB}}{\rm{yr}} \,.
\end{equation}
The detection rate for the low frequency becomes
\begin{eqnarray} \label{eq: low freq all rate}
 R_{{\rm{LF,\, radio \, GRB}}} &=& 41.4 \cdot 0.866 \cdot \frac{\Omega_{\rm{LF}}}{4 \pi - \Omega_{\rm{s,\,LF}}} \, \frac{\rm{GRB}}{\rm{yr}} = \nonumber \\
 && \\
 &=& 2.65 \cdot 10^{-1} \, \frac{\rm{GRB}}{\rm{yr}} \nonumber \,.
\end{eqnarray}
Results in Eqs. \ref{eq: high freq all rate} and \ref{eq: low freq all rate} give more chance to a GRB serendipitous detection at radio frequencies.

\section{Discussions and conclusions} \label{sec: Discussions and conclusions}
In this paper, we discussed the possibility use the SKA for GRB radio detections and surveys.
In the last part of the previous section, we calculated some serendipitous detection rates. However, to detect GRBs in radio band, it is worth  noticing  that SKA dishes will be used principally to go on-source with specific (and precise) coordinates, namely after a GRB alert. In other words, GRBs will not be detected randomly, but moving antennas on-source after a satellite previous detection (e.g., Robo-AMI, mentioned above). In our opinion, only considering a comprehensive  synergy among the radio band detections  and the detections in other wavelengths ($\gamma$-ray, X-ray, optical, infrared) will be possible to really understand the light curves of these  phenomena. In fact, the SKA will considerably be able to contribute to an increment of GRB radio data, so that the statistics will be improved, and finally the understanding of the GRB physics will be enhanced. \\
On the other hand, a serendipitous detection by the SKA could give an important contribution to the development  of GRB science. In fact, another point to highlight concerns \textit{prompt} radio emissions. Even though these emissions might not exist, it is also important to point out that  they have never been directly observed by a radio telescope. Currently, this very short emission might be detected randomly only, because moving an antenna on-source takes some minutes (or seconds, in the best case). In collaborations like  Robo-AMI and \Swift, after the satellite detections, the radio antennas go on-source in $\sim 4$ minutes but this time is too slow for a radio observation in the GRB \textit{prompt} phase. \par 
It is important to stress that GRBs are peculiar sources and  basically emit  covering all the spectrum. The principal investigations are limited around high frequencies (i.e., $\gamma$, X, optical) and often authors try to point out any single aspect of them, to classify  these objects in  several  classes. However, because of the very large emission spectrum, GRBs should be studied with a different approach, that is without focusing only on a narrow range of the electromagnetic spectrum. Analyzing GRBs with a large ``spectral field of view'' would give the chance to see aspects related to each other among more bands, perhaps discovering common features which are not clear if one sees only specific details in narrow electromagnetic bands. GRB studies in radio band could become crucial for discoveries in this sense, therefore observations by appropriate telescopes are necessary. Additional radio data will give the possibility to find possible correlations either within the radio band or between the radio frequencies and the higher ones. 
Without a lot of radio observations and focused analyses, it is impossible to get  precise and self-consistent observations. In particular, we cannot say  that every GRB with a fluence greater than an estimated value can emit in radio band (because only observations can confirm or not  this issue), but the results shown here about the radio detection rates and GRB radio sensitivities are relevant enough to encourage the study in this direction by the SKA.  \par
Another important  issue is related to   cosmology. As it is well known, GRBs are cosmological sources and their studies  can be addressed in a cosmological perspective.  The most ambitious goal is demonstrating that  these objects could be  used as \textit{distance indicators}. In order to find intrinsic  relations amomg GRB parameters that effectively  translates into distances, it is worth noticing  that only observational quantities should be used. However, obtaining a standard  light curve for GRBs, as Phillips achieved for SNeIa \citep{PhillipsRelation}, is  extremely more difficult. Indeed, GRBs spread in a very large range of variables that could be used to fix their fundamental features.  If the redshift is known,  it is possible to obtain a distance  for GRBs. The SKA can help in this goal considering also   observations not directly related to GRBs (e.g., the spin-flip emission of the hydrogen at 1.4 GHz from the host galaxy). \par
To conclude, precise radio observations and surveys would allow a deeper and  complete understanding of these still mysterious objects. The expansion of the SED for GRBs would help to relate different bands to each other and a focused study also in radio frequencies would open a research perspective  which has not been explored yet in detail. In this work we have shown that this opportunity is a fact since the SKA will be able to observe and detect GRBs. It is also useful to point out   that the radio band is not affected by radiation extinction, contrary to higher frequencies. In general, an accurate calorimetry for radio well-detectable GRBs would be possible. In addition, radio observations can be useful  for  estimations of  the inverse Compton scattering, since only radio frequencies can probe the density of the interstellar medium. However, it is imperative to keep in mind that  synergies among ground-based and satellite telescopes, observing at different frequencies,   have to be realized in order  to  understand the fundamental nature of GRBs.  
%
%
\section*{Acknowledgements}
%
%
%
This research  used  data and software provided by the following institutes and databases:  the High Energy Astrophysics Science Archive Research Center (HEASARC) at the  Astrophysics Science Division,  NASA/GSFC;  the  High Energy Astrophysics Division at  the Smithsonian Astrophysical Observatory; 
the SIMBAD database at the CDS, Strasbourg, France;
 the UK \textit{Swift} Science Data Centre at the University of Leicester;
  the VizieR catalogue access tool, CDS, Strasbourg, France. The original description of the VizieR service is published in A\&AS 143, 23. \\
This research  used also  the Jochen Grainer's Table on \href{http://www.mpe.mpg.de/~jcg/grbgen.html}{http://www.mpe.mpg.de/$\scriptstyle\mathtt{\sim}$jcg/grbgen.html}. \\
The Authors acknowledge the \textit{Societ\`a Aerospaziale Mediterranea S.c.r.l.} (SAM), the \textit{European Industrial Engineering S.r.l.}  (EIE - Group) companies for technological information and the Regione Campania  (\textit{Dottorato in Azienda} project), 
the {\it Progetto R.A.D.I.O. - Radiotelescopi per Azioni di Internazionalizzazione e cooperazione},
POR CAMPANIA FESR 2007-2013 Ð O. O. 2.1.
 \\ 
The Authors acknowledge also M. G.  Dainotti, T. Di Girolamo, P. Millici  for discussions and  suggestions on the topics. 
SC acknowledges financial support of INFN ({\it iniziativa specifica} TEONGRAV). This article is also based upon work from COST action CA15117 (CANTATA), supported by COST (European Cooperation in Science and Technology).


\section*{Appendix A} \hypertarget{sec: Appendix A}{}
Table \ref{Table: all GRBs} is the merging of catalogues mentioned and explained in part in Sec. \ref{Collection of GRB catalogues}. For each row there are seven columns. The first two columns are GRB coordinates (RA J2000 and DEC J2000) in decimal degrees; the third, fourth and fifth are the acquisition time with year, month, day (days are decimal and contain information about hours, minutes and seconds); the sixth column is the referring mission and, finally, the source name. \\
In the sixth column, only one of the missions which detected  GRBs is reported. As previously mentioned, some GRBs have been detected by more than one mission, hence they could be reported in more than one catalogue. \\ 
As for source name, we  reported the same name of the catalogue considered in the sixth column. No official names were found for the BNT GRBs, therefore only ``GRB'' has been written in the last column. Furthermore, BT source names are the same used in the HEASARC web table, with a final dash instead of the standard progressive letters. However, sources have been sorted by detection time, so it is possible to read all GRBs in a time sequence. \par
The references used for every mission are reported in  Table \ref{Table: all GRBs}:
\begin{description}
 \item[- GRANAT satellite (cat. GRA)] \hfill \\
    References in the \href{http://heasarc.gsfc.nasa.gov/W3Browse/gamma-ray-bursts/phebus.html}{\textit{phebus}} catalogue \citep{GRANAT.HEASARC1994, GRANAT.HEASARC1995, GRANAT.HEASARC1998, GRANAT.HEASARC2002}. 
 \item[- BATSE/CGRO (t) (cat. BT)] \hfill \\
    references in the \href{http://heasarc.gsfc.nasa.gov/W3Browse/cgro/batsegrb.html}{\textit{batsegrb} catalogue} \citep{BATSE4B.Meegan, BATSE3B.Meegan} and \cite{BATSE.Stern}. 
 \item[- BATSE/CGRO (nt) (cat. BNT)] \hfill \\
    references in the \cite{BATSE.Stern} and \cite{BATSE.Kommers}. 
 \item[- Konus/Wind (cat. K/W)] \hfill \\
    references in the \cite{Pal'shin2013}. 
 \item[- BeppoSAX (cat. BeS)] \hfill \\
    references in the \cite{BeppoSAX}. 
 \item[- HETE-2 (cat. HET)] \hfill \\
    references in the {``\textit{hete2grb}: HETE-2 Gamma-Ray Bursts''} by \href{http://space.mit.edu/HETE/Bursts/}{MIT} (Massachusetts Institute of Technology) and \href{http://heasarc.gsfc.nasa.gov/W3Browse/hete-2/hete2gcn.html}{``\textit{hete2gcn}: HETE-2 GCN Triggers Catalog''}. 
 \item[- INTEGRAL (cat. INT)] \hfill \\
    references in the \cite{Mineav2014}, \cite{Bosnjak2014} and \cite{Mereghetti2013}. We matched with the  the \href{http://swift.gsfc.nasa.gov/archive/grb_table}{``\textit{Swift} GRB Table and Lookup''} (selecting INTEGRAL mission).  
 \item[- \textit{Swift} (cat. Swi)] \hfill \\
    references in \cite{Swift} and \href{http://swift.gsfc.nasa.gov/archive/grb_table}{``\textit{Swift} GRB Table and Lookup''}. 
 \item[- Agile (cat. Agi)] \hfill \\
    references in \cite{Galli2013}, \cite{Pal'shin2013}, \cite{Longo2012} and \cite{Hurley2013} and I matched them with \href{http://swift.gsfc.nasa.gov/archive/grb_table}{``\textit{Swift} GRB Table and Lookup''} (selecting Agile mission). 
 \item[- \textit{Fermi} (cat. Fer)] \hfill \\
    references in \cite{Fermi}, \cite{Fermi2.vonKienlin} and the \href{http://heasarc.gsfc.nasa.gov/W3Browse/fermi/fermigbrst.html}{\textit{fermigbrst} catalogue} in HEASARC archive. 
\end{description}
The following list  explains how  matchings among catalogues have been achieved. As mentioned in Sec. \ref{Collection of GRB catalogues}, different GRB catalogues, gathering in each list only a single mission (the only exception is CGRO with two different catalogues). Equal events from various tables have been made as one. For this purpose, lists have been matched each other, using the following criteria:
\begin{itemize}
 \item The Agile list is matched with all other lists. A GRB is the same if it exploded within 0.005 days with respect to another one, and if they have a difference $\leq 90^\circ$ in RA and in DEC each other. Even if the Agile angular resolution is within a few arcmin, many GRBs are localized by satellite triangulations and other satellite are not so fine in localization accuracy. As a check, a  larger angular range has been used. More details about this mission can be read at the \href{http://agile.asdc.asi.it/}{ASI web site}.
 \item The BeppoSAX list is matched with all CGRO lists, as explained previously (time delay of 0.005 and angular range $\leq 146^\circ$).
 \item Among BATSE/CGRO lists no change is adopted for the criterion used by \cite{BATSE.Stern}. 
 \item \textit{Swift} list is matched with \textit{Fermi} one, with a delay of 0.005 days and angular range of $45^\circ$.
 \item The GRANAT list is matched with all other lists, using only a time delay of 0.005 days.
 \item The HETE-2 list is matched with BeppoSAX and \textit{Swift} lists: angular range of $83^\circ$ and time delay of 0.005 days.
 \item The INTEGRAL list is matched with Hete-2, \textit{Fermi} and \textit{Swift} lists. Time delay 0.005 and angular range $90^\circ$.
 \item The Konus/Wind list is matched with all other lists, with a delay 0.005 and angular range $90^\circ$.
\end{itemize}
It is worth noticing that the RA and DEC difference thresholds depend on the largest 
pointing-range error among different satellites. For  example, by 
matching three catalogues from BeppoSAX, Swift and HETE-2, where the 
worst pointings has a range error respectively of $85^o$, $67'$ and $11.9^o$, 
the value  $85^o$ is  the threshold range. One has to  choose this value in order to have the worst case.
However, in the reported cases,  the most important threshold is the time threshold. 
The time delay of 0.005 days (that is $\sim 7$ minutes between an explosion and the next) is in common. We changed angular ranges instrument by instrument, depending on their error positions, or FoV, or angular resolution. These angular values are probably large, but a conservative case has been preferred. However, the most relevant matching filter was the time delay and, that other criteria could be used.

\section*{Appendix B} \hypertarget{sec: Appendix B}{}
GRBs are observed in different bands and there is a preferred measure of unit for each band. A measure can usually be given either in luminosity $L$ (\Luminosity), fluence $S$ (\Fluence), flux $F$ (\Flux) or flux density $F_\nu$ (W m$^{-2}$ Hz$^{-1}$). Since we want to compare different wavelengths to each other, we must have the same measure unit. For convenience, the relations among there units is written here: 
\begin{eqnarray}
 L &=& \frac{4 \pi F_\nu d_L^2}{1 + z}  \\
 \nonumber \\
 F &=& F_\nu \, \Delta \nu   \\
 \nonumber \\
 S &=& F \, \Delta t \,, 
\end{eqnarray}
where $\Delta \nu$ and $\Delta t$ are respectively the bandwidth and the integration time during the acquisition. \\ 
In order to plot the SED of the 95 GRBs detected by \cite{Chandra2012}, we choose to convert everything into fluxes. Taking into account tables 1 and 4 in that work, from the fluence at 15-150 keV we can obtain the flux by dividing by the $T_{90}$. Furthermore, from data in flux density, we can calculate the flux of the source knowing the bandwidth (100 MHz at radio band and R-band\footnote{Its Full Width Half Maximum (bandwidth $\Delta \lambda$) is 138 nm, its Effective Wavelength Midpoint $\lambda_{\rm{eff}}$ for standard filter is at 658 nm \citep{GalacticAstronomy}.} at optical wavelengths). The conversion from energies into frequencies, or wavelengths, is given by using the Planck constant
\begin{equation}
 E = h \nu = h c \lambda \,.
\end{equation}
Finally, the flux density is expressed in $\mu$Jy, so, in cgs system,  it is 
\begin{equation}
 1 \, \mu\rm{Jy} = 10^{-6} \, \rm{Jy} = 10^{-32} \frac{\rm{W}}{\rm{m}^2 \, \rm{Hz}}  = 10^{-29} \frac{ \rm{erg} }{ \rm{s} \, \rm{cm}^2 \, \rm{Hz}} \,.
\end{equation}
In practice, if one has the R-band flux density $F_\nu = 6.5$ $\mu$Jy, it has be multiplied by the bandwidth 138 nm (converted into cgs system and frequency, it is $138 \cdot 10^{-7}$ cm $\times 3 \cdot 10^{10}$ cm/s) and the conversion factor is $10^{-29}$ erg s$^{-1}$ cm$^{-2}$ Hz$^{-1}$. Thus the result in flux is $F = 2.691 \cdot 10^{-23}$ \Flux.

\end{document}